\documentclass[sigconf]{acmart}

\AtBeginDocument{%
  }

\copyrightyear{2025}
\acmYear{2025}
\setcopyright{cc}
\setcctype{by}
\acmConference[CHI '25]{CHI Conference on Human Factors in Computing Systems}{April 26-May 1, 2025}{Yokohama, Japan}
\acmBooktitle{CHI Conference on Human Factors in Computing Systems (CHI '25), April 26-May 1, 2025, Yokohama, Japan}\acmDOI{10.1145/3706598.3713103}
\acmISBN{979-8-4007-1394-1/25/04}

\usepackage{enumitem}
\newcounter{goalCounter}

\usepackage{xr}

\begin{document}

\title{Divisi: Interactive Search and Visualization for Scalable Exploratory Subgroup Analysis}

\author{Venkatesh Sivaraman}
\email{venkats@cmu.edu}
\orcid{0000-0002-6965-3961}
\affiliation{%
  \institution{Carnegie Mellon University}
  \city{Pittsburgh}
  \state{Pennsylvania}
  \country{USA}
}

\author{Zexuan Li}
\email{zexuanli@umich.edu}
\affiliation{%
  \institution{University of Michigan}
  \city{Ann Arbor}
  \state{Michigan}
  \country{USA}
}

\author{Adam Perer}
\email{adamperer@cmu.edu}
\affiliation{%
  \institution{Carnegie Mellon University}
  \city{Pittsburgh}
  \state{Pennsylvania}
  \country{USA}
}

\renewcommand{\shortauthors}{Sivaraman et al.}

\begin{abstract}
Analyzing data subgroups is a common data science task to build intuition about a dataset and identify areas to improve model performance. 
However, subgroup analysis is prohibitively difficult in datasets with many features, and existing tools limit unexpected discoveries by relying on user-defined or static subgroups.
We propose exploratory subgroup analysis as a set of tasks in which practitioners discover, evaluate, and curate interesting subgroups to build understanding about datasets and models. 
To support these tasks we introduce Divisi, an interactive notebook-based tool underpinned by a fast approximate subgroup discovery algorithm. 
Divisi's interface allows data scientists to interactively re-rank and refine subgroups and to visualize their overlap and coverage in the novel Subgroup Map.
Through a think-aloud study with 13 practitioners, we find that Divisi can help uncover surprising patterns in data features and their interactions, and that it encourages more thorough exploration of subtypes in complex data.
% Analyzing subgroups in a dataset that have different outcomes than average is an important way to build intuition about data. However, subgroup analysis is difficult when the dataset is large, contains many features, or requires intersecting multiple features together to find meaningful subgroups. Automatic subgroup discovery is a promising strategy to mine interesting groups of instances from large datasets, but existing approaches take too long to run interactively, do not extend to multiple outcomes of interest, and produce a static list of results. We propose Divisi, an interactive search-based tool for data scientists to find and curate subgroups of interest within large datasets with discrete features. Divisi incorporates a novel approximate subgroup discovery algorithm that allows balancing between multiple ranking functions, and it performs faster than existing approaches while maintaining high accuracy compared to exhaustive search. The computational notebook-based visual interface provides interactive re-ranking, editing, and curation features including a novel Subgroup Map visualization. Through a think-aloud study with 13 data scientists, we find that Divisi enables analyses that would have previously been prohibitively time-consuming, and that it facilitates an understanding of the overlap and coverage of subgroups of interest.
\end{abstract}



\begin{teaserfigure}
  \includegraphics[width=\textwidth, alt={Screenshot of Divisi with six callouts for key features, including running a subgroup discovery algorithm, interactively re-ranking subgroups, comparing metrics, editing subgroup definitions, visualizing overlap and coverage, and saving subgroups for review. The left side of the interface lists ranking functions: error coverage, error rate, simple rule, and subgroup size. The middle shows the Subgroups Table, where subgroups include Inflight wifi service = "neutral", Seat comfort = "neutral", Class = "business", and six others. Two subgroups are selected and plotted in the Subgroup Map at right, showing that they have a small degree of overlap but are mostly distinct.}]{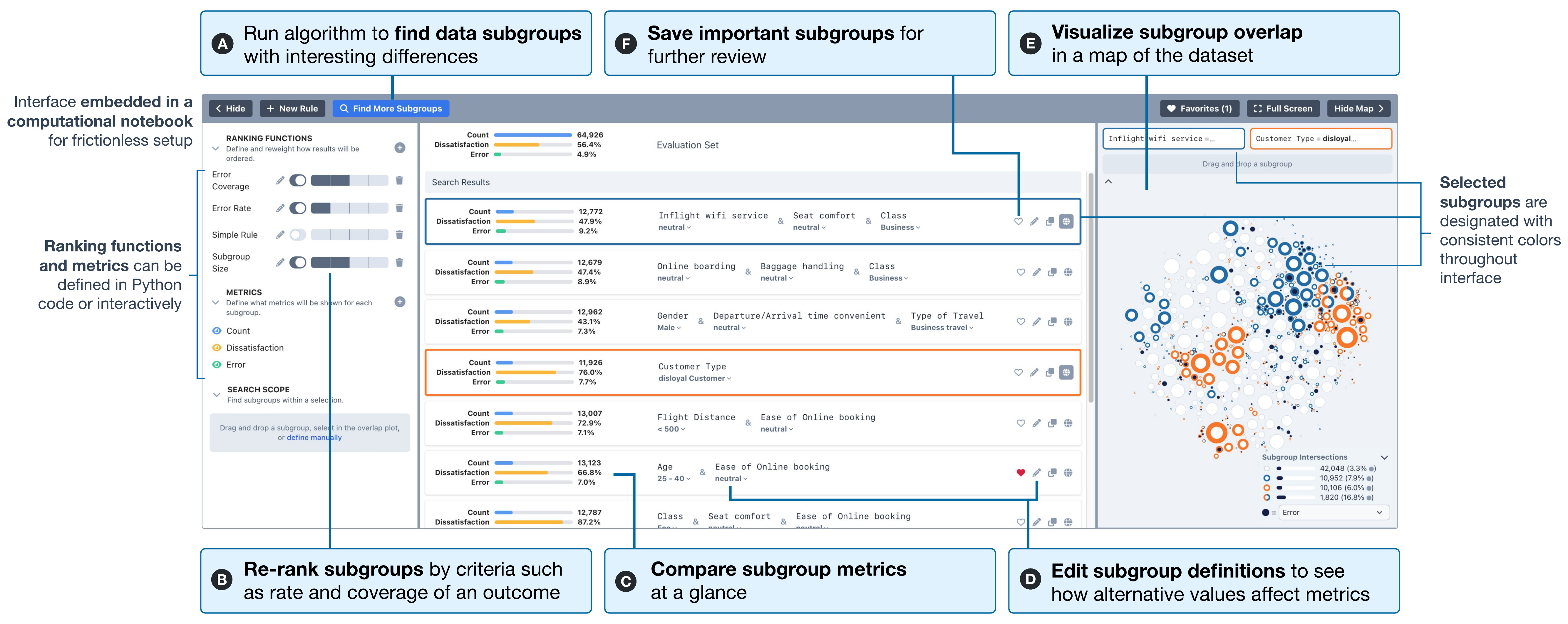}
  \caption{Divisi is an interactive visualization system to help data scientists perform \textit{exploratory subgroup analysis} on large datasets with many feature dimensions, such as the dataset of airline passenger satisfaction ratings shown~\cite{noauthor_airline_2019}. Implemented as a computational notebook widget, Divisi includes a novel approximate subgroup discovery algorithm (A) which allows interactive re-ranking by customizable functions, such as error rate and coverage (B). Users can compare metrics across subgroups (C) and test alternative rule definitions (D) to evaluate subgroups. Finally, the Subgroup Map (E) depicts overlap and coverage between groups, so users can curate the most representative subgroups for review (F).}
  \Description{Enjoying the baseball game from the third-base
  seats. Ichiro Suzuki preparing to bat.}
  \label{fig:teaser}
\end{teaserfigure}


\maketitle

\section{Introduction}

Core to data science work is making sense of large datasets to understand what patterns they contain and how they can be used~\cite{cao_data_2018}.
Along with the ``datafication'' of our personal, commercial, and medical lives, datasets aim to capture increasingly diverse and complete representations of the world~\cite{lycett_datafication_2013}.
To build a robust understanding of these datasets, data scientists and analysts often break the data into meaningful \textit{subgroups}, or smaller subsets of the data, and characterize how these subgroups differ from the overall population.
For example, businesses use customer segmentation, a form of subgroup analysis, to understand what types of customers are most likely to purchase a product or respond to a marketing campaign~\cite{noauthor_customer_2024}.
In medicine, researchers gain insight into clinical trial results by searching for subpopulations (e.g., men over 65 with kidney failure) where a treatment had greater benefits or risks~\cite{wang_statistics_2007}.

However, many questions that could be answered using subgroup analysis remain challenging to explore, particularly in the larger, higher-dimensional datasets used in modern machine learning~\cite{helal_subgroup_2016}:
What kinds of customers most often buy products or cancel subscriptions based on their purchase history?
What combinations of prescriptions result in patients getting more frequently admitted to the hospital?
What kinds of large language model (LLM) prompts are more often correctly answered by one model versus another?
Instances in these cases could have hundreds or thousands of data features, representing diverse subtypes that are hard to isolate using handcrafted subgroups.
% We present new algorithmic pattern mining techniques, and new interactive data science workflows to interpret their results, that can help build a complete understanding of these datasets.
% Advances in algorithmic pattern mining techniques, and new interactive data science workflows to interpret their results, are needed to build a complete understanding of these datasets.
Comprehensively answering such questions requires new algorithmic pattern mining techniques that are designed to work in tandem with interactive data science workflows.

Most existing approaches for subgroup analysis either depend on the user to define subgroups to analyze~\cite{cabrera_zeno_2023,cabrera_fairvis_2019,wu_errudite_2020}, or they return a static list of groups (often termed ``data slices'') that the user must interpret ~\cite{chung_slice_2020,pastor_looking_2021,eyuboglu_domino_2022}.
For complex datasets, subgroup analysis requires a combination of the two approaches to leverage both expert knowledge and algorithmic scalability.
However, the interactions in prior visual analytics tools for subgroup analysis~\cite{suresh_kaleidoscope_2023,zhang_sliceteller_2022,kwon_rmexplorer_2022,cabrera_zeno_2023} focus on refining individual slices rather than helping the user curate a more diverse set of candidates.
As a result, these tools are useful in identifying and validating patterns that already fit within an expert's mental model, but they are less effective at cultivating a more holistic understanding of the data through unexpected discoveries. % when posing the question ``what kinds of instances lead to an outcome,'' 

We propose to conceptualize subgroup analysis not as a hypothesis testing or machine learning problem, as has been done in prior work~\cite{chung_slice_2020,eyuboglu_domino_2022}, but rather as part of exploratory data analysis (EDA).
Data scientists use EDA to summarize the main characteristics of a dataset and build an intuitive understanding of its contents as they prepare it for modeling or other downstream purposes~\cite{tukey_exploratory_1970}.
However, traditional EDA techniques focus on understanding one feature at a time or interactions between two features across the dataset, rather than characterizing subgroups~\cite{wongsuphasawat_voyager_2016,epperson_dead_2023,stolte_2002_polaris}.
We define \textit{exploratory subgroup analysis} as a process analogous to EDA in which data scientists can discover, evaluate, and curate interesting and meaningful subsets of a dataset. This framing encompasses existing techniques such as rule mining~\cite{zhang_sliceteller_2022}, cluster interpretation~\cite{Cavallo2019}, and iterative subgroup refinement~\cite{slyman_vlslice_2023}, and creates opportunities for new interactions.
Exploratory subgroup analysis complements existing EDA techniques by providing an intermediate level between the overall dataset, which is often too complex to understand in its entirety, and individual instances, which can be time-consuming and misleading to interpret.

To materialize our proposed workflow for exploratory subgroup analysis and assess how it might influence data scientists' workflows, we developed a subgroup discovery algorithm and computational notebook-based visualization tool called Divisi.
The algorithm underlying Divisi expands on previous slice discovery approaches to support sense-making in interactive data science by (1) allowing for multiple arbitrary subgroup metrics and (2) using approximation to scale more efficiently to large, wide datasets.
These advances enable novel interactions in the Divisi interface, such as dynamic re-ranking and targeting subgroup search to a user-defined subset of the data.
The system also provides a novel Subgroup Map visualization which helps users relate subgroups to the dataset's overall structure.
As we illustrate through a performance evaluation and a use case with large language model (LLM) prompts that provoke unsafe responses, Divisi's algorithm and interactive workflow can reveal patterns in real-world datasets in a lightweight, efficient manner.

We explored how Divisi could help perform exploratory subgroup analysis tasks in a think-aloud study with 13 experienced data scientists.
Our results showed that data scientists particularly value the \textit{discovery} of unexpectedly important variables in the rules surfaced by Divisi's algorithm.
They heavily utilized subgroup editing and refinement tools to \textit{evaluate} whether a group was meaningful and how each feature in the rule contributed to its deviation from average.
Participants also found ways to \textit{curate} the subgroups they found for stakeholders, ensuring that their selections represented distinct subpopulations and covered most of the interesting outcomes.
Finally, they envisioned ways to fit the various features of Divisi into their projects, suggesting opportunities for future tools to more thoroughly and rigorously support exploratory subgroup analysis.

This work makes the following contributions to the literature on interactive tools for data science:
\begin{enumerate}
    \item \textbf{A proposed workflow and design goals for exploratory subgroup analysis} informed by interviews with three experts in subgroup analysis;
    \item \textbf{A novel subgroup discovery algorithm} designed specifically to support the needs of interactive applications, such as configuring the level of approximation and weighing multiple metrics of interest;
    \item \textbf{An interactive system called Divisi} that incorporates tools to tailor the discovery process, evaluate subgroup rules, and visualize their overlap and coverage; and
    \item \textbf{Results from a think-aloud study} with 13 data scientists, yielding implications for future exploratory subgroup analysis tools.
\end{enumerate}
\section{Related Work}

Our work is informed by foundational paradigms in visual analytics including exploratory data analysis and exploratory search (Sec. \ref{sec:related-eda}). 
We also build on many prior methods for subgroup analysis from data mining and machine learning, the design space of which we describe in Sec. \ref{sec:related-subgroup-analysis}.

\subsection{Exploratory Data Analysis and Search}
\label{sec:related-eda}

\citeauthor{tukey_exploratory_1970} describes \textit{exploratory data analysis} (EDA) as ``looking at data to see what it seems to say''~\cite{tukey_exploratory_1970}.
EDA is therefore distinct from hypothesis testing, or confirmatory data analysis, in its emphasis on generating insight from the \textit{data} rather than prior knowledge and expectations.
Many systems for EDA are informed by interaction techniques for \textit{exploratory search}, in which people navigate through and query information resources to build understanding about some latent concept of interest~\cite{white_exploratory_2009}.
In these interactive search settings, features such as sorting, filtering, and faceted searches~\cite{yee_2003_faceted} play a key role in helping users uncover useful information.
Applied to EDA, these techniques can enable steerable recommendations of how to visualize data features~\cite{wongsuphasawat_voyager_2016,lee_2021_lux} or efficient overviews of text data~\cite{felix_texttile_2017}.
We draw inspiration from these search techniques in the design of Divisi.

A wide variety of EDA techniques have been developed for different types of data, including small-scale tabular settings~\cite{wongsuphasawat_voyager_2016,lee_2021_lux}, high-dimensional data~\cite{Liu2017}, text data~\cite{felix_texttile_2017}, and general unstructured data~\cite{Smilkov2016}.
It is often easiest to find useful insights in EDA on tabular data because the features are generally intrinsically interpretable. 
In contrast, for text or image data the ``features'' (words or pixels) may not have any meaning on their own, making it difficult to interpret what instances have in common.
As datasets grow larger, there may also be many different subtypes within the dataset, limiting the insight provided by top-level metrics and distributions.
For this reason, many prior works aim to mitigate the complexity of large, high-dimensional datasets by automatically deriving semantically meaningful features or ``concepts'' to bootstrap the analysis process~\cite{suresh_kaleidoscope_2023,kim_interpretability_2018}.
Alternatively, some systems allow the user to define their own features of interest~\cite{wu_errudite_2020,cabrera_zeno_2023}.
However, these methods require the user to already know roughly what concept they are looking for, limiting their opportunities to explore and find unexpected patterns.
Our work relies on the presence of interpretable tabular features for every instance; however, we design for use cases in which the data scientist wants to find the relevant features out of a large set of potentially-meaningful set of descriptors.
This can afford the simplicity of working with tabular data while not restricting the analysis to the user's prior hypotheses.

% \begin{enumerate}
%     \item EDA \cite{tukey_exploratory_1970}, exploratory search \cite{white_exploratory_2009,marchionini_exploratory_2006} - what are the activities involved in each?
%     \item Faceted browsing~\cite{yee_2003_faceted}, sort and filter
%     \item More modern notions of EDA: text exploration, image exploration, embedding analysis
%     \item Benefits of traditional EDA
%     \item Challenges in extending the traditional notions of EDA to modern, large-scale datasets: multiple driving phenomena or subtypes, many variables (possibly more than can be reasoned about), uninterpretable variables
% \end{enumerate}

\subsection{Tools for Subgroup Analysis}
\label{sec:related-subgroup-analysis}

Sometimes called slice discovery, cluster analysis, or rule mining, subgroup analysis is an important part of data science that can help people understand phenomena in a dataset~\cite{liu_exploratory_2020,gamberger_active_2003}, help model builders diagnose and fix issues~\cite{piorkowski_aimee_2023,zhang_drml_2022,cabrera2021deblinder,robertson_angler_2023,zhang_sliceteller_2022, jain_distilling_2022}, explain model predictions~\cite{ribeiro_anchors_2018}, or even be used in place of a model~\cite{lavrac_decision_2004}.
However, it is usually all but impossible to define clear-cut, interpretable subgroups that exactly capture the outcome of interest (e.g., model errors), creating a design space of trade-offs for how to produce useful insights.
A wide array of subgroup analysis techniques have been developed, varying across several dimensions:

\textit{Conceptualization of a subgroup.} Differences in data types, user needs, and algorithm formulations give rise to different definitions of what a subgroup is. 
At the most subjective level, subgroups can be any semantic human-readable description of instances, regardless of whether it is encoded in the data, such as ``images of people with glasses''~\cite{cabrera2021deblinder}. 
They can also be defined by numerical proximity to some conceptual entity, such as a direction or neighborhood around an instance in an embedding space~\cite{eyuboglu_domino_2022,kim_interpretability_2018,ahn_escape_2023}. 
Finally, subgroups can be defined more precisely by constructing rules for membership, such as textual patterns~\cite{wu_errudite_2020,robertson_angler_2023} or predicates on tabular features~\cite{kwon_rmexplorer_2022,hurley_interactive_2022}. 
While the latter results in the clearest subgroup definitions, it also requires crafting or mining high-quality rules.

\textit{Source of initiative.} Many subgroup discovery methods require the data scientist to define subgroups themselves, using the affordances of the various subgroup concepts described above~\cite{cabrera_zeno_2023,wu_errudite_2020,kwon_rmexplorer_2022}. 
These methods are flexible and often provide useful insights on known areas of interest, but it can be difficult to find \textit{new} subgroups without spending time perusing individual instances. 
Algorithm-initiated approaches can provide a strong initial set of subgroups to explore~\cite{chung_slice_2020,zhang_sliceteller_2022}; however, these techniques heavily focus on producing the most relevant set of subgroups in the initial query.
There is currently a lack of \textit{mixed-initiative} subgroup analysis approaches that allow the user to interactively steer the algorithm's output to produce more relevant slices.
When subgroup analysis tools do offer mixed-initiative interactions, it is typically to \textit{refine} the subgroup definitions~\cite{slyman_vlslice_2023} or to characterize and assess their validity~\cite{hurley_interactive_2022}, both of which are supported in Divisi within our broader mixed-initiative workflow.

\textit{Visualization.} Designs to visualize and compare subgroup-level data characteristics are largely dependent on the way the subgroups are conceptualized.
For example, most clustering-based tools use dimensionality reduction scatter plots, which provide a valuable overview of the dataset but are difficult to map to data features~\cite{Liu2019,slyman_vlslice_2023,xuan_attributionscanner_2024,suresh_kaleidoscope_2023,sivaraman_emblaze_2022}.
For handcrafted subgroups on tabular data, brushable histograms can serve as controls to define predicates that are then visualized in strip plots~\cite{cabrera_fairvis_2019} or domain-specific visualizations~\cite{kwon_rmexplorer_2022}.
To visualize rule-based subgroups generated by an algorithm, table representations with sparkline charts or glyphs are often preferred as they can efficiently present summary statistics over many subgroups~\cite{kahng_visual_2016,kerrigan_slicelens_2023,zhang_sliceteller_2022}.
Similarly, UpSet plots~\cite{2014_infovis_upset} provide a dense visual representation of metrics within multiple set intersections.
Divisi combines several of these elements, including the scatter plot and the subgroup table with sparklines, with novel adaptations for tasks such as assessing overlap and coverage.

\textit{Algorithmic approach.} We can divide prior algorithms for subgroup discovery into four broad classes: lattice search, frequent itemsets, classification, and clustering.
Lattice search methods, such as Slice Finder~\cite{chung_slice_2020,sagadeeva_sliceline_2021}, \textsc{Premise}~\cite{hedderich_label-descriptive_2022}, and the Nugget Browser~\cite{guo_nugget_2011}, perform combinatorial search of a space of discrete rules to find those that most satisfy the algorithm's desirability criteria.
These methods can result in easily-interpretable subgroups, but they tend to scale poorly to datasets with hundreds or thousands of features due to combinatorial explosion.
Frequent itemset-based methods, such as DivExplorer~\cite{pastor_looking_2021} and the method developed by \citeauthor{suzuki_rule_2023}~\cite{suzuki_rule_2023}, draw on efficient algorithms from data mining such as FPgrowth, then score and rank the returned subgroups.
Similarly, these methods work best with a relatively small number of possible feature combinations.
Classification-based methods can overcome some of the performance considerations of lattice search and frequent itemset approaches \cite{yuan_isea_2022,yuan_visual_2022}, but their results often require significant work to interpret.
Finally, clustering-based methods aim to group together instances by similarity in a high-dimensional space such as a learned embedding~\cite{zhang_manifold_2019,eyuboglu_domino_2022,kim_interpretability_2018}.
Though these methods can provide insight into unstructured data, they often require a trained model, sometimes one that is jointly trained with natural-language representations, limiting their applicability.
Moreover, like classification methods, the resulting clusters and concepts are not always straightforward to interpret because of their reliance on learned embeddings.
Divisi builds on this extensive space of previous algorithms, adopting a modified lattice search approach that addresses scalability issues using approximation.
While it is most directly applicable to tabular datasets as a result, we propose ways to use it in unstructured data contexts in Sec. \ref{sec:use-case}.

Because there are so many alternative techniques for subgroup analysis, each with their own specific associated data types and challenges, there is not a clear consensus of what approach should be applied to a given problem.
As a result, data scientists may not typically include subgroup analysis in the exploratory phase of their workflows.
Our work aims to make it easier to perform subgroup analyses interactively within a typical programming environment, and we assess in our study whether they might find such capabilities useful in their daily work.

\section{Formative Design: Exploratory Subgroup Analysis Workflow}
\label{sec:formative}
% We developed a novel slice discovery algorithm called Divisi that extends this previous work by addressing the following design requirements:
% \begin{enumerate}[label={\bfseries G\arabic*.}, ref={\bfseries G\arabic*},itemsep=1ex]
%     \item \textit{Flexible slicing by multiple weighted criteria.} Errors are often just one of many phenomena data scientists seek to understand about the models they develop and their underlying data~\cite{holstein_improving_2019}. Rather than searching only for slices by a single binary metric, our approach was designed to support ranking by several metrics at once (e.g., positive true labels and negative predictions). \label{goal:sd-score-fns}
%     \item \textit{Configurable to prefer faster performance or more thorough search.} Existing slicing algorithms~\cite{sagadeeva_sliceline_2021,pastor_looking_2021} typically perform exhaustive search over slices that meet the minimum support constraints, which can be prohibitively slow for large, wide datasets. We aimed to allow users to view \textit{approximations} of the optimal set of slices at customizable levels of speed and coverage. \label{goal:sd-approximate}
% \end{enumerate}

To establish the design goals for our system, we adapted \citeauthor{pirolli_sensemaking_2005}'s sense-making framework~\cite{pirolli_sensemaking_2005} with insights gained from semi-structured interviews with three data scientists experienced in subgroup analysis.
Sense-making captures how analysts move between individual observations and larger-scale, more rigorous hypotheses, similar to the process of insight discovery in EDA~\cite{tukey_exploratory_1970,cabrera_what_2022}. 
However, it is unclear how the stages of sense-making might correspond to exploratory analysis steps on \textit{subgroup} data.
As such, we aimed to understand how data scientists who regularly work with subgroups interpret and reason about this type of data and translate their specific observations to high-level assessments.

We recruited three data scientists with extensive experience in subgroup analysis in different domains, including health technology (participant E1), algorithmic fairness (E2), and biomedical informatics (E3).
We asked these practitioners about the current role that subgroup analysis plays in their workflow when analyzing data or building models, and about the challenges they face when analyzing subgroups (see Sec. B.1 in the Supplementary Material for the questions used).
We also gave them open-ended interpretation tasks using example subgroups in three different interfaces, one consisting of plain text and two early variants of the Divisi interface.
Through these tasks we sought to understand how they would make sense of these subgroups and what they thought of the system design.
Each interview took place remotely, lasted around 60 minutes and was recorded and transcribed for analysis.

The practitioners we interviewed used subgroups for a variety of reasons, ranging from evaluations of model performance to understanding patient subpopulations with different treatment needs in medical data.
For example, E1 described using subgroup analysis after building a model for screening clinical trial participants, with the goal of determining what kinds of instances led to more predicted exclusions.
To validate that the model was not discriminating by any factors such as race, gender, or socioeconomic status, they explored subgroups sliced by each factor individually as well as intersections of these factors.

Experts particularly valued subgroups defined by clear-cut rules as a way to communicate model behavior and dataset patterns to less technical stakeholders.
However, consistent with the lack of a dominant subgroup analysis method in the literature, all three experts reported primarily relying on prior or external knowledge and ad-hoc visualizations to perform subgroup analysis.
They faced challenges in deciding which features to use for defining subgroups, assessing whether the differences they saw were real, making sense of interactions between features, and comparing metrics across multiple subgroups.

\begin{figure}
    \centering
    \includegraphics[width=\linewidth, alt={Flowchart showing four stages, Dataset, Subgroups, Schema, and Synthesis, connected by three cyclic paths labeled Discover, Evaluate, and Curate. The Discover cycle consists of (1) Find Subgroups going from Dataset to Subgroups, and (2) Investigate Data Features going back to Dataset. The Evaluate cycle consists of (3) Schematize going from Subgroups to Schema, and (4) Search for Evidence going back to Subgroups. The Curate cycle consists of (5) Build Case going from Schema to Synthesis, and (6) Find Gaps going back to Schema.}]{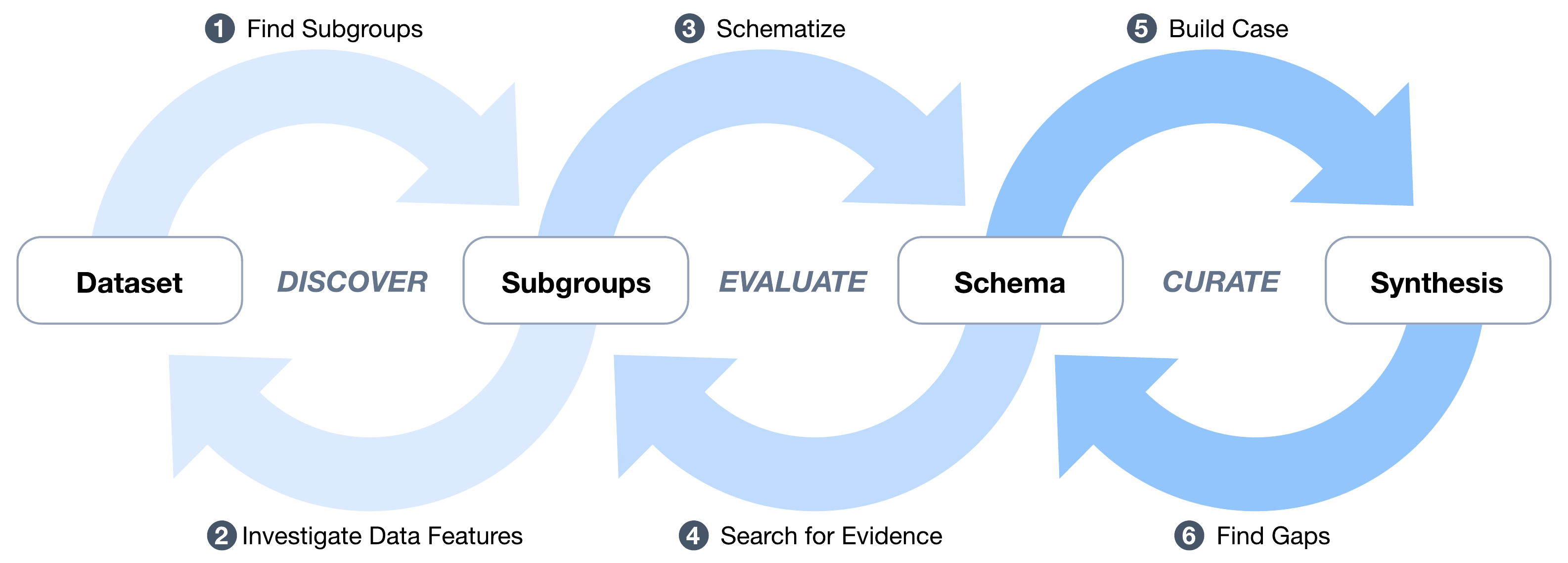}
    \caption{Proposed workflow for exploratory subgroup analysis, adapted from \citeauthor{pirolli_sensemaking_2005}'s sense-making framework~\cite{pirolli_sensemaking_2005} and informed by three expert interviews.}
    \label{fig:esa-model}
\end{figure}

Based on the practices and challenges identified in these interviews, we condensed the relevant parts of the \citeauthor{pirolli_sensemaking_2005}  framework into four successively higher-level representations: the original data, sets of subgroups, schemas or mental models of behavior, and a synthesized understanding of the dataset.
In the resulting workflow, shown in Fig. \ref{fig:esa-model}, the analyst moves between these representations through three iterative activities: discovery, evaluation, and curation.
We note that this adaptation of the sense-making process is similar to other process models proposed in visual analytics, such as workflows for exploratory analysis of machine learning models by \citeauthor{cabrera_what_2022}~\cite{cabrera_what_2022} and \citeauthor{zhang_manifold_2019}~\cite{zhang_manifold_2019}.
Though the overall analysis goals are consistent with prior work, specialized algorithmic and interactive tools may be required to conduct this sense-making process on data subgroups, informing the design of Divisi.
Below we describe how the activities of discovery, evaluation, and curation map to tasks requiring special affordances for subgroup analysis, following \citeauthor{munzner_nested_2009}'s taxonomy of visual analytics design~\cite{munzner_nested_2009}.

% Key findings from the formative study: practitioners value subgroup analysis because it provides intuition about the data, and because subgroup analyses can easily be interpreted by less technical stakeholders.
% However, people most often just use off-the-shelf tools for subgroup analysis and prior knowledge to decide what to slice by.
% They find it challenging to do subgroup analysis in large datasets with many features, and they would like to explore possible ways to pre-process the data before slicing.

% \begin{enumerate}
%     \item Automatic subgroup discovery to improve rigor and help sort through many features
%     \item Balance multiple metrics of interest.
%     \item Support reasoning about interactions between multiple features.
%     \item Use subgroups to help characterize the makeup of the dataset.
% \end{enumerate}

\textbf{Discovery.} Identifying important subgroups to look at was a major challenge for the experts we interviewed, particularly E1 and E2.
For example, E2 recounted a project about measuring disparities in medical record data:
\begin{quote}
\textit{``There's, I don't know, thousands of diagnosis codes at different levels.... So we had to do some literature review to justify why we're including those particular ones [rather] than the other ones. The ones we included are the ones that are known to [have disparities].''}
\end{quote}
% Not only was it challenging to find the initial features to analyze, the experts also found it difficult to visualize and reason about interactions about features (E1, E2).
Given this complexity, participants wanted tools that automatically identify subgroups to provide a \textit{``less biased view of the data''} (E3).
At the same time, they expressed the need to incorporate their prior knowledge, and that of stakeholders, to guide subgroup definitions.

\begin{enumerate}[label={\bfseries T\arabic*.}, ref={\bfseries T\arabic*},itemsep=1ex]
    \setcounter{enumi}{\value{goalCounter}}
    \item \textit{Find Subgroups:} The system should provide fast, approximate algorithmic subgroup recommendations as a starting point for analysis. \label{task:find-subgroups}
    \item \textit{Investigate Data Features:} Users should be able to define and store custom rules to probe the effects of features they identify as interesting. \label{task:investigate-data-features}
    \setcounter{goalCounter}{\value{enumi}}
\end{enumerate}

\textbf{Evaluation.} After identifying potentially interesting subgroups, experts wanted quantitative, actionable ways to detect if the subgroup was meaningful to their analysis.
% While statistical significance was a factor in assessing subgroups, our respondents were most interested in just looking at the metrics that would be most relevant to downstream decisions: \textit{``It's usually associated more with some type of business outcome or cost metric... [such as] bad press, bad marketing, or to make sure they're doing things the right way''} (E1).
E1 described the difficulty of comparing metrics across several different subgroups to decide which were the most meaningful, while E2 emphasized that the relevant performance metrics could be different for different subgroups.
Participants also wanted to see how their metrics of interest would change as they added and removed features from the rules to assess intersectionality (E1, E2).

\begin{enumerate}[label={\bfseries T\arabic*.}, ref={\bfseries T\arabic*},itemsep=1ex]
    \setcounter{enumi}{\value{goalCounter}}
    \item \textit{Schematize:} The system should enable comparison of the metrics in and overlap between multiple subgroups. \label{task:schematize}
    \item \textit{Search for Evidence:} Users should be able to create variants of subgroups that reveal how individual features contribute to the metrics and interact with each other. \label{task:search-evidence}
    % \item \textit{Visualize overlap between subgroups and their coverage of the dataset.}
    \setcounter{goalCounter}{\value{enumi}}
\end{enumerate}

\textbf{Curation.} The final product in all three experts' subgroup analyses was a presentation that could allow stakeholders to interpret the subgroups and the differences in metrics associated with them.
This was particularly true for problems in which domain experts were needed to interpret the subgroups: \textit{``In our analyses we usually... present the information and then let [the medical expert] make decisions based on their results''} (E3).
To make sure they were conveying accurate information to these non-data-scientist stakeholders, participants expressed the desire to make sure they captured an overall sense of the dataset during their exploration (E1, E2, E3).
They also suggested that the subgroups they identified could be used to familiarize others with the data provided they had attained good coverage.
\begin{enumerate}[label={\bfseries T\arabic*.}, ref={\bfseries T\arabic*},itemsep=1ex]
    \setcounter{enumi}{\value{goalCounter}}
    \item \textit{Build Case:} Users should be able to build a collection of subgroups that can be presented to stakeholders.\label{task:build-case}
    \item \textit{Find Gaps:} The system should surface under-explored areas of the dataset for further schema development and refinement of the search criteria.\label{task:find-gaps}
\end{enumerate}
\section{Subgroup Discovery Algorithm}
\label{sec:algorithm}

% \item \textit{Provide ways for users to configure algorithmic results for faster performance or more thorough search.} Existing slice discovery algorithms~\cite{sagadeeva_sliceline_2021,pastor_looking_2021} typically perform exhaustive search over subgroups that meet the minimum size constraints, which can be prohibitively slow for large, wide datasets. We aimed to allow users to view \textit{approximations} of the optimal set of subgroups at customizable levels of speed and coverage. \label{goal:sd-approximate}
    % \item \textit{Support balancing multiple criteria of interest.}
Existing subgroup discovery algorithms (as reviewed in Sec. \ref{sec:related-subgroup-analysis}) are predominantly designed to be run once and to retrieve subgroups that are mathematically optimal by some predefined criterion.
While these approaches are useful when the analyst has a specific goal in mind, their running time and lack of flexibility suggest they may be less compatible with the interactive exploratory subgroup analysis process described above.
We therefore developed a novel algorithm that better supports this iterative sense-making process by addressing two key requirements implied by the design goals in Sec. \ref{sec:formative}: 

\begin{itemize}
    \item \textit{Configurably approximate search.} Scalability is a major consideration for exploratory subgroup analysis, as data scientists may want to perform subgroup discovery many times as they refine their goals and intents. To support task \ref{task:find-subgroups} (Find Subgroups), the algorithm should allow users to control how deeply to search for subgroups, allowing them to get initial results quickly.
    \item \textit{Defining and weighing multiple criteria of interest.} Existing algorithms~\cite{chung_slice_2020,pastor_looking_2021} primarily focus on a single criterion (such as error rate), and they find problematic subgroups relative to the entire dataset. However, task \ref{task:find-gaps} (Find Gaps) requires data scientists to change their search criteria on-the-fly or target the search to specific regions of the dataset. Allowing users to search by multiple criteria at once could help them more flexibly express what constitutes an ``interesting'' subgroup.
\end{itemize}

Below we describe how our algorithm meets these requirements, and we present an evaluation of its accuracy and performance on three datasets compared to prior slice discovery approaches.
Throughout these sections, we use the running example of the UCI Census Income dataset~\cite{adult_2}, which consists of records for 48,842 individuals with 12 distinct features.
We train a classification model to predict whether each individual has an income of at least \$50K/yr, resulting in an error rate of 11.5\%.
Our goal in the running example is to find meaningful subsets of the data in which the model errs more often than average.

\subsection{Problem Setup}

Intuitively, the objective of subgroup discovery is to find subsets of a dataset, each defined using a clear-cut rule, that have interesting differences in some metric compared to the overall dataset.
For example, in the Census Income dataset, we could define a subgroup using the rule \texttt{relationship = "Husband" \& age = "45 - 65"}.
We can formalize this task by defining some notation: Let $X$ be a matrix of values representing a dataset, with $N$ instances and $M$ features per instance.
Note that each feature has to be discrete or categorical in order to define rules, so numerical features need to be binned into discrete categories.
Our goal is to find subgroups $S$, where each group is defined by a rule combining up to $L$ features in the form \texttt{X1 = "v1" \& X2 = "v2" \& ...}.
We then score and order these subgroups according to \textit{ranking functions}, which are each defined as functions of some length-$N$ outcome vector $Y$ and the subgroup $S$.

The simplest way to achieve the above, i.e. generate the top rule-based subgroups for any given set of ranking functions, would be to exhaustively score and rank all possible intersections of up to $L$ features.
However, even for a relatively small dataset with 100 features and two values per feature, there are over 1.3 million rules possible with $L = 3$. % (200+200*198/2+200*198*196/6)
Divisi therefore aims to ensure that the subgroups it returns are close to those that would be returned by naive iteration, but \textit{without} enumerating all possible feature combinations.

% We note that this formulation of subgroup discovery is similar to previous slice-based evaluation techniques such as Slice Finder and SliceTeller~\cite{chung_slice_2020,zhang_sliceteller_2022}, with the addition of arbitrary user-defined ranking functions.
% Unlike other approaches such as Domino and the Spotlight~\cite{eyuboglu_domino_2022,deon_spotlight_2022}, Divisi does not depend on learned representations for each instance, so in order to work with unstructured data it requires tabular metadata features to be extracted for each instance.
% However, it is often more straightforward to generate a large set of discrete-valued features to describe unstructured data, than it is to identify which features are most discriminative for subgroup analysis.
% We demonstrate the feasibility of this approach using the Reviews dataset in Sec. \ref{sec:performance-eval}, and we discuss further applications in the Discussion (Sec. \ref{sec:discussion}).

% As with other slice discovery methods, Divisi requires that the input features be discrete or categorical. Users can specify the list of variables to be used for slicing through an external configuration in Tempo query language syntax, either by adapting the variables used for modeling with the \tqlinline{cut} command or by defining new variables altogether. In our running example, health data analyst Ava uses the binary and categorical variables they had previously defined as-is, and they discretize the continuous variables into ``low,'' ``normal,'' and ``high'' bins.

\subsection{Sampling Approach}
\label{sec:sampling-approach}

Given the discrete-valued input matrix described above, Divisi works by sampling a small number of data points, then performing approximate subgroup discovery constrained to rules that \textit{match} the sampled point. 
In other words, the results of the subgroup discovery algorithm will consist exclusively of groups that contain at least one of the sampled points. 
Our key insight is that as more rows are sampled, the likelihood of finding any subgroup that matches a reasonable proportion of the dataset increases \textit{independently} of the size of the dataset. 
Users can therefore configure how many points to sample (and by extension, the running time of the algorithm) based on the size of subgroup they are looking for.
Furthermore, by sampling specifically from rows that contain outcomes of interest (such as positive labels or errors), we can further reduce the computation needed to find the most relevant groups.

\begin{figure*}
    \centering
    \includegraphics[width=0.75\linewidth, alt={Diagram showing how the Divisi sampling approach works on a toy dataset with five columns X1-5, and an outcome metric Y. The fourth row is sampled, with values X1 = 0, X2 = 1, X3 = 1, X4 = 0, X5 = 0, and Y = 0. From this, five rules are scored, one for each feature value. The top two rules are X2 = 1 and X4 = 0, and these are expanded in the next stage by testing adding an additional feature to them. This process repeats until the maximum rule length is reached.}]{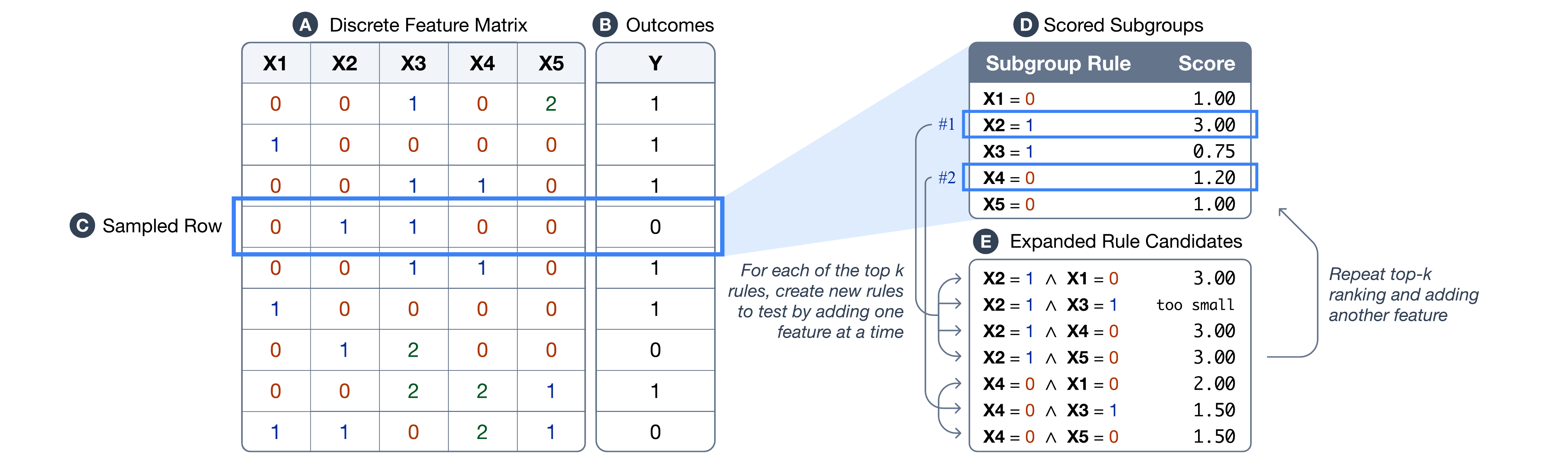}
    \caption{Divisi's subgroup discovery algorithm takes as input a matrix of discrete-valued input features (A) and one or more score functions, in this case a Binary Outcome Rate score over the outcomes in (B). For each sampled row (C), the algorithm first scores each single-feature slice containing that row (D), then iteratively expands the top $k$ slices using other features that match the sampled row (E). In this example, $k = 2$ and the minimum slice size is 2 instances.}
    \label{fig:slice-finding}
\end{figure*}

Divisi uses a beam search algorithm to progressively find high-scoring rules with more input features (illustrated in Fig. \ref{fig:slice-finding}). For each of the $n$ sampled ``source rows'' (highlighted row in Figs. \ref{fig:slice-finding}A and B), the algorithm first scores and ranks all univariate rules that contain the row according to all ranking functions separately (Fig. \ref{fig:slice-finding}C). Then, the top $k$ rules according to \textit{each} ranking function are expanded by one feature, again testing all single-feature additions that match the source row (Fig. \ref{fig:slice-finding}D). Subgroups that contain a smaller proportion of the dataset than the user-defined minimum size $p_\text{min}$ are filtered out. This process continues until the user-defined maximum number of rule features $L$ has been reached, and the algorithm returns the ranked results over all subgroups that were evaluated during any iteration. 

In summary, Divisi provides five parameters to address the goal of configurably approximate search: the sample count $n$ (default 100), a binary mask within which to sample source rows (default none), the minimum subgroup size $p_\text{min}$ (default 0.01), the beam size $k$ (default 50), and the maximum rule length $L$ (default 3).
In practice, it is sufficient to set $n$ and $p_\text{min}$ based on the size of the desired subgroups.
% \edit{In summary, Divisi provides the following parameters to address the goal of configurably approximate search, though in practice only $n$ and $p_\text{min}$ need to be set based on the user's preferences:
% \begin{itemize}
%     \item \textbf{Sample count $n$ (default 100):} the number of rows that serve as a source row for the beam search algorithm. Larger values increase running time and the likelihood of finding smaller subgroups. This is distinct from downsampling the dataset, as subgroups are still scored over the entire dataset by default. Moreover, this does not limit the number of subgroups that can be found.
%     \item \textbf{Sampling mask (default none):} a binary mask over the rows of the dataset indicating where to sample source rows from. For large datasets, sampling specifically from rows that contain the outcome of interest can reduce the number of samples needed.
%     \item \textbf{Minimum support $p_\text{min}$ (default 0.01):} the smallest proportion of the dataset that a returned subgroup can have. Larger values decrease running time by eliminating subgroups that are too small.
%     \item \textbf{Beam size $k$ (default 50):} the number of top rules that are expanded in each iteration of the beam search algorithm. Larger values increase running time by including more possible intersections.
%     \item \textbf{Maximum rule length $L$ (default 3)}: the maximum number of features that can be included in a rule. Larger values increase running time by evaluating more complex rules.
% \end{itemize}
% }

\subsubsection{Testing Robustness} \label{sec:testing-robustness}
Prior approaches for slice discovery typically either do not account for the robustness of subgroups or they use statistical methods such as $\alpha$-investing to mitigate the false discovery rate problem~\cite{chung_slice_2020,pastor_looking_2021}. Assessing the reliability of a subgroup is important for Divisi, particularly when the group is small relative to the overall dataset; however, $p$-value thresholds may be less appropriate for exploratory analysis across several different metrics. Therefore, we split the data into \textit{discovery} and \textit{evaluation} sets~\cite{green_subgroup_2021}, such that all initial scoring and ranking occurs on the discovery set while interactive re-ranking utilizes the evaluation set. This ensures that the metrics displayed for each subgroup are obtained separately from those used to develop the results.

\subsection{Ranking Functions}

Ranking functions compute a non-negative value for each subgroup that is higher if the subgroup more closely matches a criterion of interest. Divisi can in theory support any function that can be implemented in Python and that operates on an outcome variable and a binary subgroup mask.
In the interactive interface, the user can select from the six pre-defined ranking functions below. Through experimentation on datasets with binary, categorical, and continuous target variables, we found that these functions covered the majority of use cases:
\begin{itemize}[leftmargin=*]
    \item \textit{Binary Outcome Rate (Precision).} This function simply measures the ratio of the mean of a particular outcome variable within the subgroup to the overall mean. More formally, given an outcome vector $Y$ and a subgroup $S$ containing a subset of instances, the outcome rate score is calculated as
    \begin{equation}
    \text{Score}(S; Y) = \frac{\sum_{i \in S} y_i}{|S|}
    \end{equation}
    This ranking function can be used to rank subgroups by the rate of any binary metric, such as positive labels, positive predictions, model errors, or even similarity to another subgroup. By taking the inverse of the function, we can find subgroups with lower rates than average.

    \item \textit{Binary Outcome Coverage (Recall).} Complementary to the outcome rate, outcome coverage measures the proportion of all instances with a positive outcome that are captured within a subgroup, also known as recall:
    \begin{equation}
        \text{Score}(S; Y) = \frac{\sum_{i \in S} y_i}{\sum_{i = 1}^{N} y_i}
    \end{equation}
    % Similar to the above, the coverage metric can be used with any binary metric. Although the definitions are equivalent, we refer to these functions as ``rate'' and ``coverage'' to avoid confusion with precision and recall as these typically relate to a specific set of ground-truth labels.

    \item \textit{Interaction Effect.} This function penalizes rules with extraneous slicing features by taking the ratio of the binary outcome rate in the current subgroup against the maximum rate in all subgroups defined by subsets of the current group's slicing features. For example, for the rule \texttt{A = 1 \& B = 3}, this function would divide the outcome rate by the maximum of the rates for the rules \texttt{A = 1}, \texttt{B = 3}, and the overall dataset. A value greater than 1 indicates that all features play a role in elevating the outcome rate.
    
    \item \textit{Mean Difference.} This ranking function measures the difference between the mean of a continuous outcome in a subgroup and the overall mean:
    \begin{equation}
        \text{Score}(S; Y) = \left|\frac{\sum_{i \in S} y_i}{|S|} - \frac{\sum_{i=1}^{N} y_i}{N}\right|
    \end{equation}
    The Mean Difference score can be used to rank subgroups in regression models by how different the in-group predictions are from the true values.

    \item \textit{Group Size.} To prioritize groups that have good support while not being too large, we score subgroups using a Gaussian function of the group size with a configurable ideal mean and spread.

    \item \textit{Simple Rule.} This function penalizes subgroups that are defined using too many features in the rule, using an inverse logarithmic function.
\end{itemize}
% Note that all of these functions are non-negative by definition, guaranteeing that the top $k$ slices for any weighted sum of the score functions will be among the top $k$ slices for each individual score function. This enables us to efficiently re-rank slices for different combinations of score functions, addressing \ref{goal:sd-score-fns}.

\subsection{Accuracy and Performance Evaluation}
\label{sec:performance-eval}

\begin{figure*}
    \centering
    \includegraphics[width=0.75\textwidth, alt={Six performance evaluation charts, three for running time in seconds and three for recall, within which each shows results for Census Income, Airline, and Reviews. Divisi was tested with 10, 20, 50, 100, and 200 samples taken, and compared to Lattice Search and Frequent Itemset implementations. In all results, running time decreases as the minimum subgroup size increases from 1\% to 10\%, and recall increases. Recall is always 100\% for Lattice Search and Frequent Itemset as these methods are not approximate, and it is 0.5 or above for all settings of Divisi. In Census Income, all running times are within 0.3-3 seconds. In Airline, Lattice Search performs comparably to Divisi (around 10 seconds) while Frequent Itemset ranges from 15 to 100 seconds. In Reviews, Lattice Search ranges from 100 to 4,000 seconds, while Divisi always remains under 300 seconds (Frequent Itemset fails to run).}]{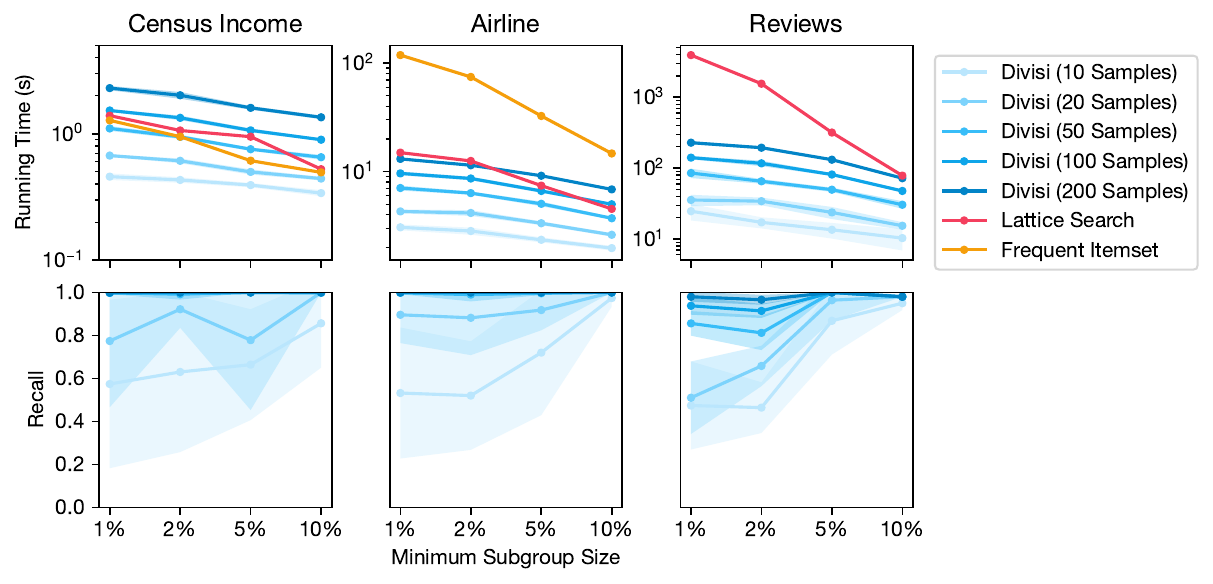}
    \caption{Average running times and accuracy (recall in top 50 returned results) for different parameter settings of Divisi, compared against a Lattice Search and Frequent Itemset approach. (We were unable to run the Frequent Itemset approach on the Reviews dataset due to excessive memory consumption, so we only report its performance on the Census Income and Airline datasets.) Shaded regions represent one standard deviation over 10 trials.}
    \label{fig:performance-eval}
\end{figure*}

To test the Divisi algorithm's ability to return results with configurable levels of approximation and speed, we conducted an evaluation of its accuracy and time performance under different parameter settings.
We performed subgroup discovery on three datasets of varying size and dimensionality: UCI Census Income~\cite{adult_2} (48K instances, 12 features), an Airline Passenger Satisfaction dataset sourced from Kaggle~\cite{noauthor_airline_2019} (129K instances, 22 features), and a subset of the Yelp review dataset\footnote{To produce a discrete-valued tabular representation of the review dataset, we used a grouped bag-of-words approach, similar to the analysis shown in Sec. \ref{sec:use-case}.}~\cite{zhang_2015_yelp} (200K instances, 2,000 features).
Further details on the datasets used are provided in Sec. A of the Supplementary Material.
% The size of the datasets ranged from 48,000 in Census Income to over 200,000 instances in Reviews, and from 12 features in Census Income to 2,000 in Reviews.

For comparison, we tested two baseline approaches: a Lattice Search algorithm similar to Slice Finder~\cite{chung_slice_2020} and a Frequent Itemset approach based on DivExplorer~\cite{pastor_looking_2021}. 
To ensure that each method would rank subgroups the same way, we used their underlying algorithms to retrieve candidate subgroups, then used Divisi's ranking functions (a Binary Outcome Rate score for error and a Group Size function) to rank the top 100 subgroups.
We measured the recall of each Divisi output against the outputs of Lattice Search and Frequent Itemset, since these two approaches are both exhaustive and produce identical results.

As shown in Fig. \ref{fig:performance-eval}, different parameter settings of Divisi can result in either more accurate or faster results. Divisi's running time (upper panels) is mostly dependent on the size of the dataset and the number of sampled rows $n$. 
In contrast, the Lattice Search and Frequent Itemset methods are markedly slower for smaller minimum subgroup sizes, which increase the number of viable subgroups. 
Divisi's recall (lower panels) generally increases as the minimum subgroup size increases and as the number of sampled instances increases, yielding a more extensive search.

Notably, Divisi remains feasible to run as data dimensionality grows regardless of parameter setting.
Its running time is comparable to the Lattice Search algorithm in the Census Income and Airline datasets, but up to two orders of magnitude faster in the wider Reviews dataset.
Moreover, Divisi can be parallelized to sample different subsets of the data in different worker threads, leading to even faster performance.
These results suggest that Divisi could be used in place of existing Lattice Search methods without sacrificing runtime or accuracy in smaller data settings, while making subgroup analysis more feasible in larger datasets.
Additionally, while rule-based subgroup discovery has been largely infeasible on large text datasets using existing methods, Divisi's efficiency makes this approach much more practical, even with thousands of word- or concept-level features.
\section{Visualization Interface}
\begin{figure*}
    \centering
    \includegraphics[width=\textwidth, alt={Screenshot of the Configuration sidebar and Subgroups Table in Divisi loaded on the Census Income dataset. A dropdown is open under a "marital-status" feature to edit the allowed values for that feature in one subgroup.}]{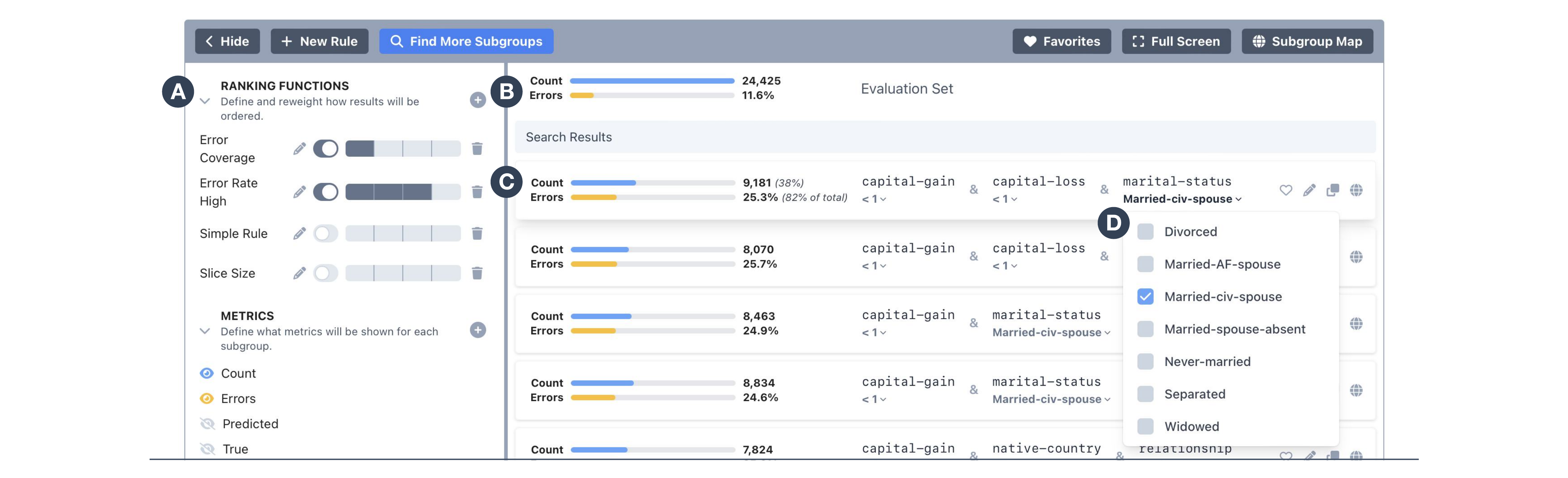}
    \caption{The Configuration sidebar (A) and the Subgroups Table (B) allow users to run the subgroup discovery algorithm and browse the rules it returns. For example, in the Census Income dataset, the first returned subgroup (C) represents people with no capital gains or losses who are married to a civilian spouse. This subgroup comprises 38\% of the dataset, and has an error rate of 25.3\%, compared to 11.6\% in the overall Evaluation Set. By clicking the dropdown next to the \texttt{marital-status} feature (D), we can test alternative values for that feature.}
    \label{fig:divisi-interface}
\end{figure*}

Integrating the algorithm described above, we developed the Divisi visualization system to support the exploratory subgroup analysis workflow synthesized in Sec. \ref{sec:formative}.
The system is designed to be installed as a Python package and run in a Jupyter Notebook environment, which is widely used in data science workflows~\cite{shen_interactive_2014}.

\begin{figure*}
    \centering
    \includegraphics[width=\textwidth, alt={Three screenshots of the Subgroup Map shown on the Census Income dataset in different states. In (A), no subgroup is selected, and shaded bubbles spread around the chart can be seen to represent model errors. In (B), three subgroups are selected, and the groups overlap such that the first overlaps with the second and third, but the second and third do not overlap with each other. In (C), a subgroup is hovered in the Subgroups Table and all bubbles are grayed out except those that match the hovered group.}]{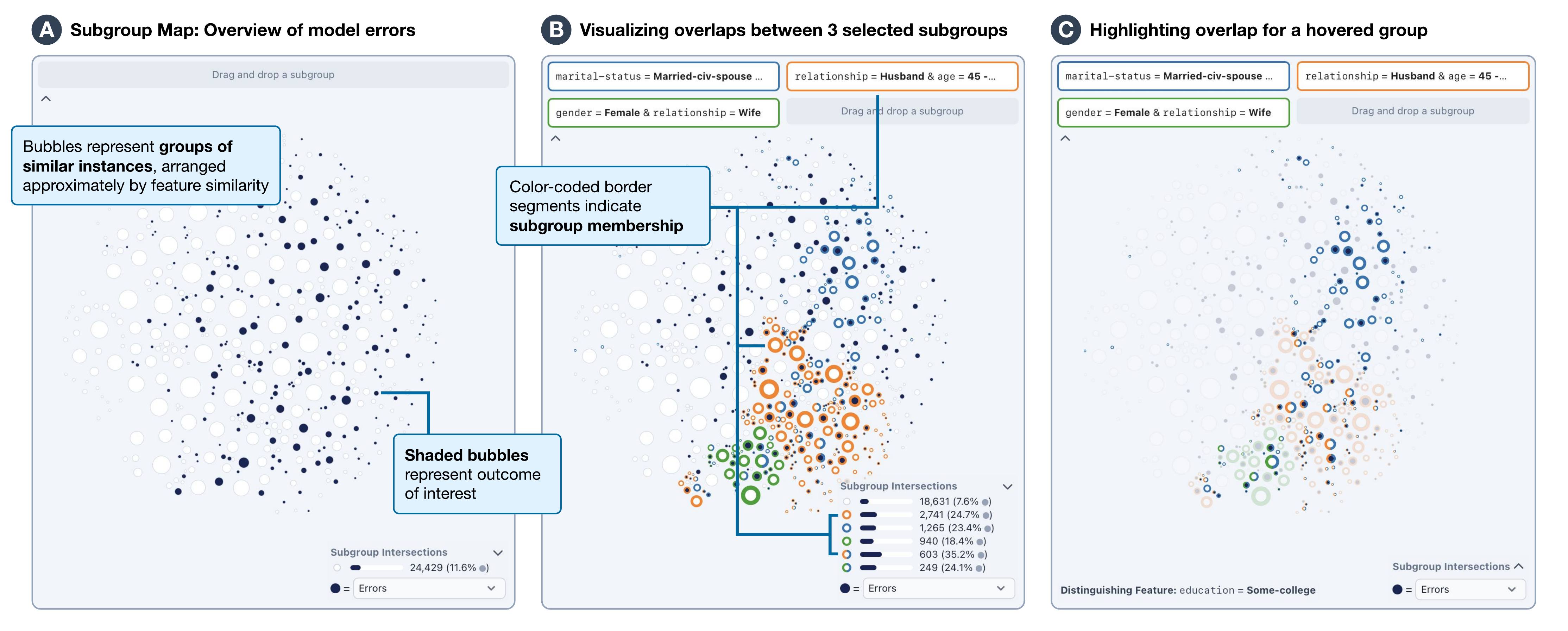}
    \caption{Different states of the Subgroup Map on the UCI Census Income dataset~\cite{adult_2}: (A) an overview of the dataset with no subgroups selected, (B) intersections between three selected subgroups, and (C) highlighting the points that match a subgroup when hovered in the Subgroups Table. Filled-in bubbles indicate classification errors for the income prediction task; each bubble's size indicates the number of instances it contains.}
    \label{fig:subgroup-map}
\end{figure*}

To summarize how Divisi's interface might support a typical workflow, let us imagine a data scientist looking for subgroups with high prediction error from the UCI Census Income dataset~\cite{adult_2}.
Upon opening Divisi in a Jupyter Notebook, the data scientist clicks \textit{Find Subgroups} to run the sampling algorithm described above. 
The results appear in the Subgroups Table (Fig. \ref{fig:divisi-interface}B), where the first subgroup consists of people who are married and have no capital gains or losses. 
The data scientist could then refine and re-rank the results using the Configuration sidebar (Fig. \ref{fig:divisi-interface}A); for example, they might upweight ``Error Rate High'' to see subgroups with higher error rates regardless of size. 
The Subgroups Table also includes controls to edit subgroups and define custom rules; for example, the user can change the value of the \texttt{marital-status} feature in a rule to see how alternative values affect the error rate (Fig. \ref{fig:divisi-interface}D).
Finally, the data scientist can drag-and-drop rules into the Subgroup Map to see how much they overlap and cover the dataset (Fig. \ref{fig:subgroup-map}).

Below we discuss in more detail how these interface components enable the three exploratory subgroup analysis activities of discovery, evaluation, and curation.
% The interface consists of three main components: the Configuration sidebar (Fig. \ref{fig:divisi-interface}A), the Subgroups Table (Fig. \ref{fig:divisi-interface}B), and the Subgroup Map (Fig. \ref{fig:subgroup-map}).
%The Configuration sidebar provides controls for the subgroup discovery process, which in turn determine the results shown in the Subgroups Table.
%Meanwhile, the Subgroup Map depicts  a bubble chart of the dataset with visual encodings to show relevant outcome metrics as well as subgroup membership.
% Users can select subgroups from the table or drag-and-drop them into the map to visualize their overlap and coverage.

\subsection{Finding and Ranking Subgroups}
\label{sec:vis-discovery}

Users initialize Divisi in a Jupyter Notebook by providing a dataset containing discrete feature values and one or more outcome metrics. 
For example, for the Census Income dataset, we could provide a binary indicator of whether each instance was mispredicted as the outcome metric, as well as the true and predicted values.
Upon launching Divisi, users can click the \textit{Find Subgroups} button to run the subgroup discovery algorithm and populate the Subgroups Table with an initial list of rules, completing task \ref{task:find-subgroups}.
Ranking functions to order the subgroups are automatically generated based on the provided outcomes, and they can be edited on the fly in the Ranking Functions section of the Configuration sidebar (Fig. \ref{fig:teaser}B).
For example, after passing in the ``Error'' metric, Divisi automatically generates a Binary Outcome Rate function to find subgroups where that metric is higher than average.

Top subgroups are computed for every ranking function provided during discovery, so the user can later sort subgroups by any weighted combination of these functions.
The Ranking Functions area allows users to toggle functions on or off, as well as to choose a weight for each function in four increments.
Subgroups are instantly re-ranked as the user adjusts the ranking function configuration, providing rapid feedback about what combination of functions leads to the most interesting results.

To perform task \ref{task:investigate-data-features} (Investigate Data Features), users can use the \textit{+ New Rule} button to define a custom subgroup.
Rules can be defined using a simple syntax, such as \texttt{"marital-status" = "Married-civ-spouse" \& "education" = "Some-college"}.
% This editing functionality also provides users the ability to experiment with ``or'' and ``not'' operations, which are not currently used in the subgroup discovery algorithm because it would drastically increase the number of possible subgroups.
Upon entering the rule, Divisi automatically computes all of the metrics for the new subgroup.

After finding one or more subgroups of interest, users can save those groups to the Favorites, which can later be viewed separately from the search results. This helps in building a case that can be presented to others (task \ref{task:build-case}).

\subsection{Assessing Feature Interactions}
\label{sec:vis-evaluation}

Addressing task \ref{task:schematize} (Schematize), the Subgroups Table lists each retrieved or custom rule along with a summary of the metrics within the subgroup represented by the rule.
Sparkline-style charts, similar to prior subgroup analysis interfaces~\cite{kahng_visual_2016,zhang_sliceteller_2022}, allow users to quickly scan over the list of subgroups and identify general patterns in the metrics as well as the surfaced data features.
% However, as with many subgroup discovery algorithms, it is important to note that multiple results could represent very similar sets of instances.
% To assess whether multiple subgroups have a high degree of overlap with one another, users can select the subgroups of interest or drag them into the Subgroup Map
This can help users combine insights across multiple subgroup results to build a higher-level understanding of one area of the dataset.

% \begin{figure}
%     \centering
%     \includegraphics[width=0.5\linewidth]{figures/slice_feature_editing.pdf}
%     \caption{The lightweight editing tools allow users to interactively test alternative values (or combinations of values) and see how their metrics change.}
%     \label{fig:slice-feature-editing}
% \end{figure}
The Subgroups Table also provides lightweight rule editing functionality to help users quickly test hypotheses about feature interactions, addressing task \ref{task:search-evidence} (Search for Evidence).
By clicking a feature's name in a rule, users can toggle that feature on and off to instantly see how the metrics change.
The feature values can also be adjusted through a dropdown menu, allowing users to select one or more alternative values as shown in Fig. \ref{fig:divisi-interface}D.
For more fine-grained changes, users can use the same syntax described in Sec. \ref{sec:vis-discovery} to edit any subgroup's definition.

\subsection{Visualizing Subgroup Overlap and Coverage}
\label{sec:vis-curation}

As with many subgroup discovery algorithms, Divisi can return subgroups that have many instances in common despite being based on different features (e.g., people who are married would likely overlap with people with a relationship type of ``husband'').
However, existing tools do not help the user assess overlap and coverage, potentially leading to analyses that focus too heavily on small areas of the data.
Divisi includes a novel Subgroup Map visualization that serves three purposes: (a) helping data scientists check whether multiple subgroups have a high degree of overlap with one another, (b) showing how much of the outcome has been covered by the selected subgroups, and (c) providing an overview of the dataset structure that can point to new areas to explore.
Early designs of Divisi used UpSet plots~\cite{2014_infovis_upset} or Venn diagrams, but in initial feedback with data scientists we found that users preferred the spatial dataset overview provided by a scatter plot, as is common in subgroup analysis tools for unstructured data~\cite{suresh_kaleidoscope_2023,robertson_angler_2023,Liu2019}.
% \footnote{Currently only binary outcomes are supported in the Subgroup Map, but the visual encodings could easily be extended to continuous outcomes in the future.}

We opted to use a dimensionality reduction plot using the t-SNE algorithm~\cite{maaten_visualizing_2008} as the starting point for the Subgroup Map.
Although dimensionality reduction can introduce distortions in proximities between points, it provides a useful way to navigate the dataset and depict groups of points that is well-established in visual analytics~\cite{Brehmer_2014_dr}.
To mitigate overdraw for large datasets, we visually simplify the plot by grouping together points that are within a threshold distance and have identical properties (e.g. same outcome value, same subgroup membership). 
This yields a bubble chart in which each bubble's size represents the count of its constituent points, its color represents the outcome value of those points, and its position is the centroid of their 2-D coordinates.
Finally, we perform a very short force-directed relaxation of the layout to remove any overlaps between bubbles, producing the starting view of the dataset as shown in Fig. \ref{fig:subgroup-map}A.

When the user selects a subgroup of interest or drags one from the Subgroups Table into the Subgroup Map, borders around the points are added to show which bubbles represent instances contained in the subgroup. 
Fig. \ref{fig:subgroup-map}B shows the Subgroup Map after selecting three groups from Census Income: married individuals with some college education (blue), married men aged 45-65 with no capital gains (orange), and married women (green).
For any bubble that contains instances in one or more subgroups, the border is divided into an equal-length arc for each subgroup that contains the bubble.
For instance, we can see that the blue subgroup intersects the orange group in Fig. \ref{fig:subgroup-map}B, based on the presence of bubbles with half-blue, half-orange borders.
Towards task \ref{task:schematize}, this visual depiction of subgroups enables users to visually map color-coded subgroups to regions of the plot and to identify when subgroups overlap with each other.

The Subgroup Map includes several secondary interactions to help users gain more information about subgroup overlap and coverage, further addressing task \ref{task:schematize}.
The Subgroup Intersections panel in the bottom-right corner of the map summarizes the size and outcome rate within each combination of subgroups (represented by a bubble glyph).
Hovering on a subgroup in the Subgroups Table grays out all bubbles on the map except those that contain instances matching the subgroup, helping users quickly check the location and overlap of a rule (Fig. \ref{fig:subgroup-map}C).
Hovering on or selecting a region of the map, meanwhile, shows a Distinguishing Feature that is most unique to that selection in the bottom-left corner of the map.

If a user finds an area of the dataset that they would like to describe using subgroups, they can lasso-select that area and conduct a targeted subgroup search, addressing task \ref{task:find-gaps}.
The Divisi algorithm enables efficient search within a selected region by simply adding another ranking function called the Selection Score, which is simply a Binary Outcome Rate function where the outcome is true if the instance is part of the selection. %\footnote{As described in Sec. \ref{sec:testing-robustness}, Divisi splits the data into a discovery and an evaluation set, and only visualizes the evaluation set to avoid generating and testing hypotheses on the same data. To build the Selection Score, we take the selected evaluation set instances and collect a similarly-sized set of their nearest neighbors in the discovery set.}
Using the Subgroup Map, the user can transition between high-level assessments of the dataset and retrieved subgroups within areas that they discover.

\subsection{Implementation Details}

Divisi is implemented as a Jupyter Notebook widget using the \texttt{anywidget}\footnote{\url{https://http://anywidget.dev}} library, with a Python backend and a Svelte\footnote{\url{https://svelte.dev}} frontend.
Visualizations are created using D3.js\footnote{\url{https://d3js.org}} for rendering and layout, and Counterpoint~\cite{sivaraman_2024_counterpoint} for state management.
Divisi is open-source and can be installed from PyPI and GitHub\footnote{\url{https://github.com/cmudig/divisi-toolkit}}.
\section{Use Case}
\label{sec:use-case}

% ** This dataset might not be the best to use for our use case because they created the dataset following 19 harm categories, which we are basically just replicating using Divisi **

To demonstrate a possible analytical process using Divisi on real data, we applied the system to the task of identifying large language model (LLM) prompts that can lead to unsafe responses.
While LLMs' emergent capabilities enable remarkable performance on many tasks, these models can also generate offensive or harmful content if not appropriately aligned with the intended values of its developers~\cite{ji_beavertails_2023}.
What kinds of prompts would result in unsafe responses? Moreover, what prompts might lead to more \textit{heterogeneity} in response safety (some responses are safe, others unsafe)? 
% We aimed to evaluate whether these questions could be answered using an exploratory subgroup analysis approach with Divisi.
To answer these questions, we analyzed the PKU-SafeRLHF dataset~\cite{ji_pku-saferlhf_2024}, which consists of about 145K LLM-generated responses to about 38K prompts. %, served as the context of our use case.
Each row contains a free-text prompt, two alternative responses to the prompt, and labels about each response's perceived safety. % assigned by a team of annotators with AI assistance.
Because the prompts were themselves created by asking various LLMs to generate requests in ``harm categories'' predefined by the dataset creators, we were able to corroborate our analysis against these categories.
However, such metadata is not required to use Divisi.

\begin{figure*}
    \centering
    \includegraphics[width=\textwidth, alt={Two screenshots of Divisi showing subgroups on the PKU-SafeRLHF dataset. In (A), the top subgroups for responses that are Both Unsafe are "launder, laundered", "poison, poisonous" and "boss, don, sir"', and "phishing, ddos, botnet.'" In (B), two subgroups are selected that have high rates of Different Safety: "address, addresses" and "try, trying". The Subgroup Map shows that the overlap of these two subgroups has an even higher rate of heterogeneity than either subgroup individually.}]{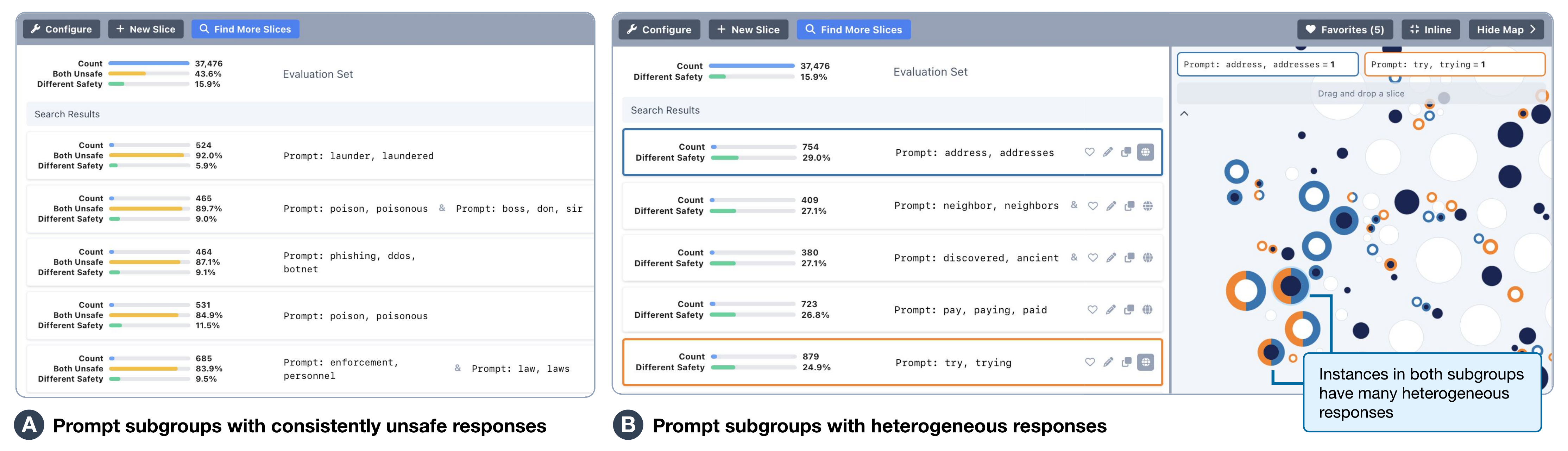}
    \caption{Subgroups identified during exploratory subgroup analysis on LLM safety evaluation data~\cite{ji_pku-saferlhf_2024}. Divisi surfaces prompts involving criminal activity as consistently unsafe (A). For prompts with more heterogeneous responses (B), one subgroup involves asking the LLM to provide people's addresses (blue group), which becomes more heterogeneous when adding the words ``try'' or ``trying'' (orange group).}
    \label{fig:llm-safety}
\end{figure*}

We used a grouped bag-of-words approach to derive metadata features for each prompt.
The unique words in the prompts were clustered using a simple word embedding model, after which we took the top 5,000 word clusters and used the presence of any word in a cluster as each metadata feature.
Because these clusters were derived from an embedding model, they represented a variety of concepts, including topics, keywords, and grammatical functions.
For instance, one feature checked whether the prompt contained cyberattack-related words (\texttt{phishing, ddos, botnet}), while another checked for quantifier words (\texttt{any, some, few, enough, least}).
For each prompt, we defined the binary output metric ``Both Unsafe'' to measure whether both candidate responses were rated unsafe, and ``Different Safety'' to capture when one response was rated safe and the other unsafe.
Running the Divisi algorithm on the 75K training data instances with 5,000 features and 200 sampled rows took roughly 6 seconds.

In the Divisi interface, we first enabled a ``Both Unsafe'' Binary Outcome Rate ranking function to find out what kinds of prompts were associated with consistently unsafe responses.
We used Divisi's Jupyter Notebook integration to pull matching instances for each returned subgroup, allowing us to peruse example prompts.
As shown in Fig. \ref{fig:llm-safety}A, many of the subgroups with the most consistently unsafe responses had to do with specific criminal activity, such as money laundering and cyber-attacks.
The fifth subgroup on the list involved an intersection of \texttt{enforcement, personnel} and \texttt{law, laws}, containing prompts about evading the authorities for which responses were both unsafe 84\% of the time.
Interestingly, by disabling the individual features one by one, we could see that the presence of \texttt{enforcement, personnel} was almost sufficient to yield consistently unsafe responses (80\%), but \texttt{law, laws} alone was only consistently unsafe 52\% of the time.
This could suggest that responses to prompts about violating laws that do not directly mention law enforcement are considered less ``unsafe.''

We also analyzed prompts in which one response was labeled safe and the other unsafe, to understand areas of greater disagreement or heterogeneity.
As shown in Fig. \ref{fig:llm-safety}B, a subgroup of prompts with requests for addresses occupied one corner of the Subgroup Map and showed a Different Safety rate of 29\%, compared to 15.9\% on average.
By hovering on each rule in the Subgroups Table, we looked for other subgroups that might cover the same area of the plot, and we found that the feature \texttt{try, trying} overlapped with the addresses bubbles substantially.
In fact, the Subgroup Intersections chart showed that for prompts that matched both rules, the rate of heterogeneity was over 38\%.
This intersection, which was small and would have been difficult to find without Divisi's visualization of overlap, could point to linguistic features that might make LLMs more likely to avoid completing the user's request.

Using Divisi, we were able to build a case comprised of several subgroups defined by interpretable, clear-cut rules, without significant manual analysis but still curated for meaningfulness.
After plotting these subgroups on the Subgroup Map, we observed that there were still many regions of heterogeneous safety that were uncovered by our selected groups.
With further analysis, we could examine this ``long tail'' of prompts by creating progressively smaller subgroups, to see if additional patterns emerge.
\section{User Study}
\label{sec:user-study}

To evaluate data scientists' use of exploratory subgroup analysis, we conducted a think-aloud study with Divisi centered around three research questions:
\begin{enumerate}
\item What is the role of subgroup analysis in data scientists’ existing workflows?
\item What insights do data scientists uncover during the different stages of exploratory subgroup analysis?
\item What opportunities do data scientists perceive to use exploratory subgroup analysis in their work?
\end{enumerate}

\subsection{Study Design}

Each think-aloud session lasted around one hour and consisted of an introduction and pre-study survey, two hands-on tasks using the Divisi interface, and a final debriefing interview (see Sec. B.2 in the Supplementary Material for the questions used). 
% We aimed to understand the current role of subgroup analysis in data scientists’ existing workflows through pre-study questions and by asking participants how they would approach the subgroup analysis tasks before using Divisi. 
% % Participants were encouraged to  throughout the analysis process.
% Then, we gathered insights into the patterns and discoveries data scientists could uncover through the two tasks. 
% Then, with the debrief questions, we gathered new opportunities participants perceived, both from their experiences with Divisi and from their own project contexts. 
% Additionally, we also evaluated how data scientists perceived Divisi’s different features and how well the features support the needs of data scientists for subgroup analysis, by gathering their feedback on the tool throughout the study session. 
%add more details of participants?
In the introduction phase, we gathered background information about the participants’ experience with data science and subgroup analysis, setting the context for the tasks that followed. Participants completed a pre-study survey that contained questions pertaining their occupation, years of experience with data and machine learning, their familiarity with Python, and their familiarity with and approach to subgroup analysis. 
%Participants were also encouraged to provide additional context verbally.

The two tasks in the study required participants to use Divisi to analyze the Airline Passenger Satisfaction dataset mentioned in Sec. \ref{sec:performance-eval}. 
The dataset contains results of a survey of 129,880 passengers’ satisfaction with the airline based on various aspects of the airline services. 
The 22 features in the dataset included demographic information (e.g., Gender and Age), traveler information (Customer Type, Type of Travel, Class), flight information (Flight Distance, Departure Delay in Minutes, Arrival Delay in Minutes), ratings for individual aspects of the flight experience (inflight Wi-Fi service, cleanliness, etc.), and an overall satisfaction rating of satisfied or not satisfied. We converted the 5-point ratings for aspects of the flight experience into ``not satisfied,'' ``neutral,'' and ``satisfied'' to simplify the space of possible subgroups for the user study.
The two tasks were as follows:
\begin{enumerate}[label={\bfseries Study Task \arabic*:}, ref={Study Task \arabic*},itemsep=1ex,labelindent=0pt, wide=0pt]
    \item \textbf{Dissatisfaction.} Participants were asked to discover and interpret subgroups of data that could provide insights into what types of customers tended to be dissatisfied with their experience. Overall, 57\% of instances in the dataset had an overall rating of dissatisfied, meaning that the outcome of interest was fairly common. We asked participants to perform any analysis they would want to be able to present a comprehensive report to stakeholders at the airline company.
    %An example of how the interface looked during this task, and the subgroup results that appeared, is shown in Fig. \ref{fig:airline-example}. 
    \label{study-task:dissatisfaction}

    \item \textbf{Model Errors.} We trained a binary classification model to predict each passenger's overall satisfaction rating, resulting in an error rate of 4.8\%. Participants were then asked to find out what subgroups tended to have higher-than-average model error rates. Similar to the dissatisfaction analysis task, participants were told to curate their insights for stakeholders of the airline company. This task was slightly more difficult because of the involvement of a classification model as well as the lower outcome base rate, which made the discovered rules slightly more complex to interpret. \label{study-task:error}
\end{enumerate}
Before the tasks began, the interviewers provided a brief tutorial of how to use Divisi using an annotated screenshot. Participants were also given guidance during the session on how to use the tool for specific tasks that they expressed interest in performing.
After completing the tasks, we asked participants to evaluate the usefulness of the various features of Divisi, and to reflect on how the system could apply to their past projects. 

We recruited 13 data scientists with some experience with machine learning and data science, particularly involving large-scale tabular or text data. 
(None of the experts from the formative work also participated in this study.)
Participants were compensated \$20 USD, and the protocol was approved by our institutional IRB.
Each one-hour session was conducted on Zoom and was recorded and transcribed for analysis.

\subsection{Analysis}
To analyze the data collected during the study, we took an open coding approach on the transcripts and video recordings of the study sessions. Two of the authors each coded all transcripts, discussing disagreements and reaching consensus as needed, to understand participants' perspectives and answer the three research questions above. The videos were used to supplement the transcripts with specifics about the participants' actions using the system. The coders generated 483 codes after annotating all of the study transcripts. We then used affinity diagramming to group codes into higher-level categories to reveal broader themes and opportunities.

\subsection{Results}
Overall, participants were able to uncover a broad range of insights into both dissatisfaction and model errors in the Airline dataset using Divisi.
They also compared the process of exploratory subgroup analysis to their current workflows, and they suggested ways to make Divisi easier to learn and more broadly applicable.

\subsubsection{Current Practices: Participants have few structured ways to perform subgroup analysis}
\label{sec:results-current-practices}
Participants in our study had varying degrees of experience with subgroup analysis (9/13 reported having used it in a prior project), but none used tools specific to subgroup analysis.
Many of those who used subgroup analysis mentioned using off-the-shelf programming libraries such as \texttt{pandas} and \texttt{matplotlib} (5/13 participants) to manually create filtering rules, which limited their opportunities for discovery: 
\begin{quote}
\textit{``When I'm doing data analysis, most of the time it's just for creating the plots using matplotlib. I'm just exploring the features that I'm interested in, but I cannot explore all of the things.''} (participant 1, denoted P1)
\end{quote}
When it came to interpreting models or identifying features associated with an outcome, there were few interactive ways to do so. 
Some participants described using clustering analyses to identify useful subgroups in an unsupervised manner (3/13).
However, these participants found clustering to be cumbersome for exploratory analysis because of the challenge of going into each cluster and interpreting what kinds of instances it contained.
Instead, most others described training simple models such as logistic regression to predict the outcome of interest, \textit{``and then just [seeing] what that spits out''} (P10).
Participants (8/13) mentioned that compared to these strategies, using Divisi would make it easier and faster to narrow down which variables to focus on:
% ``I think it'd be really helpful to see the slices that i'm not aware of'' (P8)
\textit{``Now I have a starting point. I can just start diving in''} (P6).
% ``this is the equivalent of like diving straight into the patient and being like, these people had symptoms, these people didn't'' (P6).

\subsubsection{Discovery: Participants identify surprising insights in subgroup results}
\label{sec:results-discovery}
All participants were able to quickly make assessments of what types of instances led to both dissatisfaction (\ref{study-task:dissatisfaction}) and model errors (\ref{study-task:error}), performing \ref{task:find-subgroups} (Find Subgroups) using the results in the Subgroups Table.
A strategy taken by several participants (6/13) was to glance over the table and find individual features and values that appeared frequently: \textit{``From this page overall I could see that people who are not satisfied with the Wi-Fi service have higher dissatisfaction rate''} (P7).
In this way, participants found that people with dissatisfied or neutral ratings with Wi-Fi service, gate location, and online boarding often tended to be dissatisfied overall.
Likewise, in \ref{study-task:error}, customers who were not part of the airline's loyalty program (identified by 8/13 participants) and those with several neutral ratings (7/13) were frequently found to be mispredicted.

In many cases, the patterns participants observed in the subgroup results went against their intuitions, suggesting that they might not have uncovered the same subgroups had they been guided only by their own expectations.
For instance, several users described expecting to see passengers who were dissatisfied with the flight delay (4/13 participants) or baggage handling (P5), neither of which were commonly-surfaced subgrouping features: \textit{``It is kind of surprising to me, because I think the departure delay... will be the most important thing that people will complain about''} (P1).
To verify that dissatisfaction with delays was indeed less than the other returned subgroups, P1 created a custom rule representing high departure delay and compared its metrics to the rest of the table (an instance of \ref{task:investigate-data-features}, Investigate Data Features).

\begin{figure*}
    \centering
    \includegraphics[width=\textwidth, alt={Screenshot of Divisi from the user study in which a participant has three subgroups selected and is visualizing their overlap in the Subgroup Map. Almost all shaded bubbles (Dissatisfaction instances) are covered by the selection. In the Ranking Functions, the user has both Dissatisfaction and Dissatisfaction Coverage enabled with equal weighting.}]{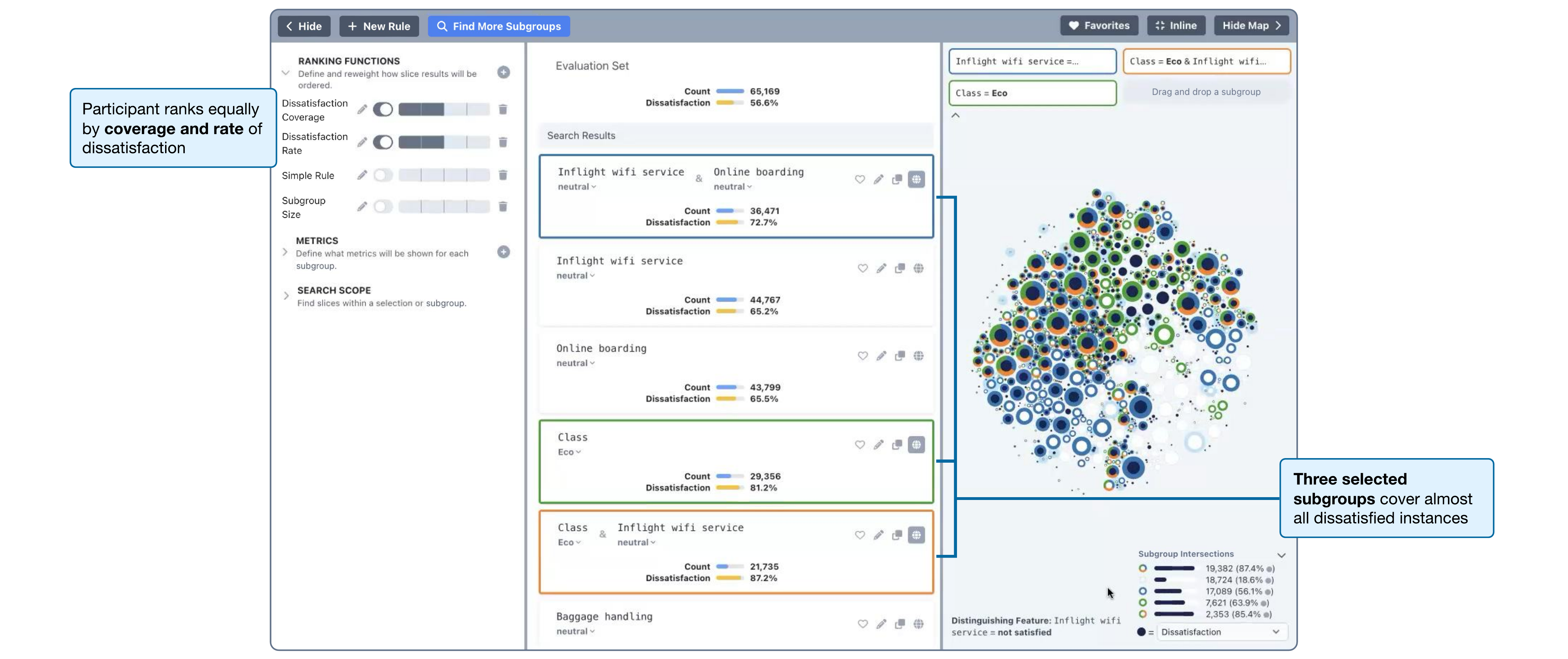}
    \caption{During \ref{study-task:dissatisfaction}, P13 investigated the intersections of three rules involving the passenger's fare class and their satisfaction with the Wi-Fi service and online boarding features. They decided to interpret these subgroups because as shown in the Subgroup Map at right, they cover almost all of the shaded (dissatisfied) bubbles.}
    \label{fig:airline-example}
\end{figure*}

Participants used the Ranking Functions extensively during the discovery process.
Although many participants (5/13) found the ranking functionality confusing at first, they ultimately found it helpful to tailor the results to the kinds of subgroups they were looking for.
The most common criterion was ranking by the Binary Outcome Rate function for dissatisfaction or error (11/13 participants); however, many (7/13) also spent time prioritizing subgroups using the Binary Outcome Coverage function.
To get a holistic sense of the dissatisfied or mispredicted instances, participants saw coverage as a way to find subgroups that were \textit{``actually most representative of that group''} (P8).
Users (6/13) also adjusted the Simple Rule function for the purposes of familiarizing themselves with the dataset's variables and finding subgroups that would be easier to explain to a stakeholder: \textit{``Whatever conclusions you have, you want them to be simple enough. People usually don't care about if you... go very deep on different clusters of customers''} (P12).
Putting these desiderata together, P11 described how they set up their ranking weights:
\begin{quote}
\textit{``First, I don't want the rules to be too complex, so I want to keep it simple, but not that simple. Not just one [feature]... And I want these rules to be representative. So actually, I want the size to be like around 10\% [of the dataset]... And also I want the percentage of the errors as high as possible.''}
\end{quote}
Since different users expressed different criteria for what a ``good'' subgroup would look like, we observed more variation in analysis paths than if only one set of rankings were available.

\subsubsection{Evaluation: Participants use subgroup editing tools extensively to test feature interactions}
\label{sec:results-evaluation}

Participants schematized their understanding of subgroup behaviors (\ref{task:schematize}) and searched for additional evidence (\ref{task:search-evidence}) in diverse ways, most often by editing existing subgroup rules and creating new ones.
Most straightforwardly, several participants tested alternative values for features to observe correlations in outcomes: \textit{``The longer the flight [the greater the satisfaction]. Your... 10,000-mile business flight has only 24\% dissatisfaction, but at 500 [miles] it's 52\%''} (P10).
By performing this lightweight editing in subgroups defined by multiple features, participants (6/13) reasoned about interaction effects, i.e., feature values that had more pronounced impacts on outcomes when a different feature took a particular value.
For instance, by interactively toggling between economy and business class customers, P12 found that \textit{``in-flight service doesn't matter as much for economy customers.''}
This type of interaction helped 3/13 participants conclude that combining ratings with different polarities could cause higher error rates: \textit{``People are maybe very satisfied with something, but they may be unsatisfied with something else''} (P1).

In addition to editing rules, 6/13 participants used the custom rule feature to test hypotheses, driven by both prior knowledge and the patterns they observed.
For example, some participants (4/13) tried to extend rules they found in the Subgroups Table with additional features:
\begin{quote}
\textit{``I think I have to look closer [at] people who were neutral on the in-flight service, and then combine more within that subgroup to see if there's something actionable... Maybe that experience intersects with something else, and those two things are intertwined and actionable to improve the service.''} (P8)
\end{quote}
Except for one experienced user who wanted to add features to a rule in bulk using Python code (P10), Divisi's rule editor was sufficient for any query that users wanted to make.
Custom rules thereby helped participants both investigate features of interest (\ref{task:investigate-data-features}) and build up a more robust understanding of feature interactions (\ref{task:schematize}).

While evaluating subgroups, some participants (5/13) wondered whether the subgrouping features could be appearing due to randomness or spurious correlations, an evaluation aspect that Divisi does not currently support.
While Divisi splits the data into discovery and evaluation sets to mitigate false discoveries (Sec. \ref{sec:testing-robustness}), it is always possible that two features are highly correlated so that slicing by one of them is sufficient to slice by both.
For example, looking at a rule with high overall dissatisfaction but a satisfied rating on the departure/arrival time, P7 expressed the desire to dig deeper into other feature values within the subgroup to test whether \textit{``people in this subgroup are not satisfied with other aspects''}.

\subsubsection{Curation: The Subgroup Map helps more experienced participants choose subgroups to prioritize.}
\label{sec:results-curation}

We observed that Curation actions, i.e. building a case (\ref{task:build-case}) and finding gaps (\ref{task:find-gaps}), were primarily done by participants with more familiarity with subgroup analysis. 
Nine out of 13 participants used the Subgroup Map and targeted search features to build understanding about the contents of the dataset and to choose which subgroups to analyze further.
Some (4 out of 9 who used the map) looked at the Distinguishing Features for each bubble to associate types of people with areas of the map: \textit{``I always expect... there's, say, a portion of the graph to be something like satisfied loyal customers.''} (P9)
Others (3/9) performed slice search within selected regions to characterize them using rules.
Furthermore, by plotting subgroups on the map and seeing how they were distributed, participants (4/9) noticed that some subgroups could be further divided into more detailed categories: \textit{``So you have like, multiple kinds of disloyal customers''} (P12).

Consistent with the Subgroup Map's goal of communicating overlap and coverage, participants (6/9) used the visualization to eliminate subgroups that covered very similar sets of instances on the plot: \textit{``They're just like, not satisfied with various aspects, but they fall in the same place''} (P12).
Others (3/9) used it to assess at-a-glance how many of the positive outcomes were covered by their selected subgroups, to decide when they had completed the task to their satisfaction.
For example, P13 began by selecting the top five subgroups, visualizing them in the Subgroup Map, then narrowing them down to capture the same population with just three groups (Fig. \ref{fig:airline-example}).
Two participants followed an iterative methodology (similar to \ref{task:find-gaps}, Find Gaps) of adding and removing subgroups from the map to cover as many shaded bubbles (positive outcomes) as possible, while conducting targeted search within the white bubbles (negative outcomes) to find subgroups to remove.
Participants who used the Subgroup Map to understand coverage found it particularly helpful:
\begin{quote}
\textit{``Without... the visualization, I would think I get the representative groups. But with it, I'm thinking, still... many errors are missing. So I think it's a good way to use the visualization, because it makes me think of if I really grasped the majority of the errors.''} (P11)
\end{quote}

% \begin{enumerate}
    % \item Participants use the subgroup map to spatialize their understanding of the different characteristics in the dataset. lasso selecting and finding subgroups within a selection ``Because they are close to each other, and I want to see like, what kind of people they will be.'' (P1) ``I didn't even realize age is a variable in this. We've not looked at age.'' (P10) ``I always expect, like, you know, there's say, a portion of the graph to be something like, you know, satisfied loyal customers, or something like that.'' (P9) ``so you have like, multiple kinds of disloyal customers'' (P12) 
    % \item Participants use the subgroup map to choose the slices they want to analyze further. `` they're just like, not satisfied with various aspects, but they fall in the same place'' (P12) ***``Because if without, without the image or the visualization, I would think I get the representative groups, but with with it, I'm thinking, still other still, many errors are missing. So I think it's a good, good way to use the visualizations, because it makes me think of if I really grasp the majority of the errors.'' (P11)
% \end{enumerate}

\subsubsection{Perceived Opportunities: Making Divisi more actionable and broadly accessible}
\label{sec:results-perceived-opps}
Participants envisioned many ways in which Divisi could support their current and prior projects, including better understanding clinical trial results (P4), analyzing test results in educational experiments (P8), and debugging image generation models using keywords in their caption text (P13).
They suggested ways to extend the exploratory subgroup analysis workflow to accommodate these use cases, such as measuring the statistical significance of subgroup differences (P7) or displaying metrics that were more relevant to the use case (P9).
Participants also wondered if Divisi's workflow could be applied to data types that were more difficult to slice by, such as images or representations of molecules (2/13).
These data types would likely require clever ways to create metadata features for rules that are both clear-cut and interpretable.

Participants considered how metadata features were defined in our user study tasks, and some wanted to adjust the feature definitions to help them mine better subgroups. 
For example, some participants (4/13) found rules with ``neutral'' rating values not very actionable, and they wanted to either exclude that feature value from subgroups or re-bin the ratings into ``satisfied'' and ``not satisfied.''
Others described wanting to remove features that they suspected would mostly contain spurious correlations, though they wanted to do so without injecting personal bias into the process: \textit{``I have my own mental heuristic, for which features... are very low signal-to-noise, and I'm trying to remove the noise''} (P6).
While participants could in theory have edited the code to create the grouping features within the Jupyter Notebook, it would be helpful to directly support this aspect of subgroup analysis within the tool.

Finally, several participants commented that the Subgroup Map (6/13) and the Ranking Functions (5/13) were initially difficult to understand.
These users suggested simplifying how subgroups were initially presented to them, such as by enabling only one ranking function at first (P4).
Those confused by the Subgroup Map commented that clearly indicating how the chart was generated would likely make it more trustworthy (P12) and clarify the relationship between bubbles and subgroups (P1).
While these features were largely perceived as useful (8/13) once participants were familiar with how they worked, more guidance could help those less experienced with subgroup analysis leverage them effectively.

\section{Discussion and Limitations}
\label{sec:discussion}

This work presented a workflow for \textit{exploratory subgroup analysis} instantiated in a novel algorithm and interactive system called Divisi.
Throughout our formative interviews and think-aloud sessions, we saw that data scientists currently take a handcrafted approach to subgroup analysis, and they want techniques to help them discover and evaluate subgroups in more principled ways.
Moreover, they want to ensure that the subgroups they analyze are representative, cover most of the outcomes of interest, and can be easily interpreted by non-technical stakeholders.
Divisi helped our study participants perform these tasks by providing starting points for each stage of their analyses, allowing them to explore ``corners'' of the dataset that would have previously been hard to anticipate.
These results, and the current limitations of our system, suggest future directions for supporting exploratory subgroup analysis.

Rule mining has long been studied~\cite{han_mining_2000} as a way to help people find patterns in data, make interpretable predictions~\cite{furnkranz_brief_2015}, and generate model explanations~\cite{ribeiro_anchors_2018}.
Divisi builds on many of these prior works, applying ideas from data mining in our construction of an approximate subgroup discovery algorithm.
However, despite the abundance of available methods to discover interpretable rules, data scientists still typically construct subgroups based on their prior knowledge and using ad-hoc processes.
We aimed to bridge this gap by framing subgroup analysis as an \textit{exploratory} task that does not require clear goals or a specific stage in the data science process.
Divisi's design proposes that exploratory subgroup analysis tools should not only support the discovery of new subgroups, but also allow for human evaluation and curation of subgroups to juxtapose algorithmic results with expert intuitions.
Participants in our study envisioned ways to integrate all three of these activities into their work, suggesting that \textbf{an exploratory approach based on our proposed workflow may make subgroup analysis more applicable and practical for today's data science practices}. %, where datasets can evolve and models are frequently updated,
% by supporting greater flexibility and interactivity, to support an exploratory workflow
% Our work builds on these prior advances
% What has changed? Data types, multiple metrics of interest, . What hasn't changed - people still need to be able to high-level describe the patterns in a dataset

Our evaluations suggest that the novel aspects of our subgroup discovery algorithm make Divisi more amenable to large datasets and support interactive sense-making.
As we showed in the performance evaluation, the ability to configure levels of approximation through Divisi's parameters could make the tool more feasible than existing algorithms to use in large, wide datasets such as text data. 
Moreover, since Divisi's algorithm made re-ranking essentially instantaneous, participants extensively tested the effects of balancing multiple ranking functions in ways that would have been time-consuming otherwise.
Many systems developed in HCI and data visualization research have required new computational formalisms~\cite{suresh_kaleidoscope_2023,robertson_angler_2023} or algorithms~\cite{perer_frequence_2014,lam_concept_2024} to enable new insights into data or machine learning models.
Similarly, our evaluation of Divisi underscores how \textbf{formulating data mining algorithms specifically for the requirements of \textit{interactive} data science workflows can facilitate more thorough exploration.}

One current limitation of Divisi's curation features surfaced by several participants was deciding how to parameterize the space of possible subgroups.
As described in Sec. \ref{sec:results-perceived-opps}, participants pointed out that which variables were included and how they were transformed into discrete values (e.g. binning 5-point ratings into ``not satisfied,'' ``neutral,'' and ``satisfied'') could have a large impact on what subgroups were identified.
Similarly, when discovering subgroups in text data as we demonstrated in Sec. \ref{sec:use-case}, the choice and clustering of words used for grouping can greatly impact the meaningfulness of the returned rules.
Defining subgroups using syntactic features, topic models, or LLM-based concept extraction techniques could all provide different lenses on the same data.
\textbf{Exploring how to help data scientists create and test feature spaces for subgroups is an important direction for future work.}
% Participants' attentiveness to how the subgrouping space was parameterized could suggest that when using Divisi, the bulk of the analytical effort essentially shifted from mining and interpreting clusters to deciding how to \textit{express} their data using interpretable features.
For example, it is likely that these interpretable features could be constructed so that they generalize across many datasets, or generated automatically using language models to kickstart analysis. %, similar to how LIWC dictionaries can be used to measure psychological characteristics in many kinds of texts~\cite{tausczik_psychological_2010}.
% Such metadata could also potentially be automatically extracted for other types of data, such as images or graphs, further broadening the applicability of rule-based subgroup analysis.
% Future work can explore how to design these ``subgrouping vocabularies'' to kickstart the exploratory subgroup analysis process on new datasets.

When integrating exploratory subgroup analysis into a data science workflow, it is important to consider how the subgroups will be used downstream, both for statistical and, more broadly, ethical reasons. 
While subgroup or slice discovery in machine learning is most often framed in terms of identifying semantically meaningful model errors, users in our studies saw Divisi as potentially useful for a wider array of exploratory data analysis tasks.
For example, participants imagined using it to dig deeper into analyses of clinical trial results to see what kinds of patients received the most benefit.
However, these analyses come with the caveat that testing too many subgroups can lead to false positive observations, particularly in smaller datasets~\cite{wang_statistics_2007}.
\textbf{Making clear distinctions about when subgroup analysis is to be used for exploratory and confirmatory purposes~\cite{hullman_designing_2021}, and constraining the space of analytical actions for each case, could mitigate subgroup analyses that lead to spurious interpretations.}

We also note that defining black-and-white rules to categorize and separate subsets of data can be at odds with the nature of the data, particularly when it represents humans.
Just as prior HCI critiques have pointed out that most ``useful'' data constructs are convenient approximations of the true concept of interest~\cite{tal_target_2023}, exploratory subgroup analysis requires an understanding that any rule-defined subgroup is an approximation of the true subpopulation.
For example, discretizing medical lab values into low, normal, and high categories elides the reality that many people's normal lab values could be considered extreme relative to the population~\cite{cohen_personalized_2021}, leading to measurement errors when defining subgroups by any fixed cutoff.
\textbf{Future research could design ways to support reflection on where rule-based subgroups might not fully capture the intended population due to nuances in the data.}
% We also note that rules to categorize and separate subsets of data can have important implications when used to inform decisions, particularly when the data represent humans.
% Just as predictive models can implicitly create black-and-white categories to drive decisions about individuals~\cite{alkhatib_street-level_2019}, decisions driven by subgroup-level insights 

Although our evaluation allowed us to observe diverse ways that data scientists could use Divisi, we note the limitation that our participants only used the tool for about one hour, with an environment and dataset we set up. 
We chose this method to ensure consistency across the tasks performed, and to minimize the time burden on participants.
Although participants were able to conduct multiple analytical tasks during their sessions, spending more time with the system and using it on their own data could provide a richer picture of how participants might use it in the long term.
Additionally, because there is no commonly-used alternative workflow that data scientists use for subgroup analysis, we did not quantitatively assess the impact of using Divisi with respect to a baseline system.
Future work is needed to validate that the findings from exploratory subgroup analysis lead to more effective models or improved dataset understanding.

% \begin{enumerate}
% \item Using Divisi on unstructured data. For these types of tasks, Divisi's approach essentially shifts the analytical effort from interpreting clusters to selecting and curating from a large set of inherently interpretable features.
%     \item Deciding how to parameterize the slicing space
%     \item In fact, extensive subgroup analysis is sometimes discouraged in these settings because with small dataset sizes, testing too many subgroups can lead to false positive observations~\cite{wang_statistics_2007}.
% \end{enumerate}
\section{Conclusion}

Subgroup analysis is an important, yet under-utilized tool in data science.
Our results suggest that combining algorithm-generated, rule-based insights with human intuition and experimentation in an interactive workflow can help practitioners develop a thorough understanding of complex datasets.
By implementing these interactions in a lightweight notebook-based tool, we hope to lower the barrier for data scientists to try subgroup discovery and to curate unexpected, interesting subpopulations in their data.
Divisi is available as an open-source package so that data scientists and HCI researchers can build on this work, helping to make exploratory subgroup analysis more feasible for a wider range of contexts.

\begin{acks}
We thank the many data scientists and machine learning practitioners who participated in our user studies and shared their perspectives with us.
We also thank Will Epperson, Katelyn Morrison, and Dominik Moritz for providing feedback on the manuscript.
This work was supported by a National Science Foundation Graduate Research Fellowship (DGE2140739) and by the Carnegie Mellon University Center of Machine Learning and Health.
\end{acks}

\bibliographystyle{ACM-Reference-Format}
\bibliography{references}

\appendix
\section{Accuracy and Performance Evaluation Details}
\label{app:perf-evaluation-details}
Table \ref{tab:performance-eval-datasets} provides an overview of the datasets used in the performance and accuracy evaluation in Section \ref{sec:performance-eval}.
The default values were used for the beam size ($k = 50$) and rule length ($L = 3$) in all runs of the Divisi algorithm, while $n$ and $p_\text{min}$ were varied as part of the experiment.
For the source mask, we targeted row sampling to rows that have a positive value for the target outcomes specified in the table.

\begin{table*} \small
\begin{tabular}{rp{3.2cm}p{3.2cm}p{3.2cm}}
\toprule
\textbf{Dataset} & \textbf{Census Income} & \textbf{Airline} & \textbf{Reviews} \\ \midrule
\# \textbf{Instances} & 48,842 & 129,880 & 211,443 \\
\# \textbf{Features} & 12 & 22 & 2,000 \\
\textbf{Target Outcome} & Income classification error & Overall dissatisfaction & Error in predicted star rating (difference of $\geq 3$ stars) \\
\textbf{Base Rate} & 11.5\% & 56.6\% & 0.63\% \\
\midrule
\textbf{Example Subgroup} & Relationship = ``husband'' \& Education = ``assoc-voc'' \& Capital Gain = ``\textless 1'' & Gate location = ``neutral'' \& Inflight wifi service = ``not satisfied'' \& Ease of online booking = ``not satisfied'' & (back, follow, following) \& (years, year) \& (said, told, asked, thought, knew) \\
\textbf{Subgroup Size} & 1.6\% & 5.7\% & 1.1\% \\
\textbf{Outcome Rate in Subgroup} & 31.6\% & 100\% & 5.1\% \\
\bottomrule
\end{tabular}
\caption{Datasets used to evaluate the running time and accuracy of Divisi, and examples of subgroups returned by the Divisi algorithm for each task.}
\label{tab:performance-eval-datasets}
\end{table*}

\section{User Study Details}

This appendix presents condensed study protocols used for the formative design and the user study. The formative design (Sec. \ref{sec:formative}) involved semi-structured interviews with three data scientists experienced in subgroup analysis to inform tool development. The user study (Sec. \ref{sec:user-study}) consisted of a think-aloud session using the Divisi tool, designed to test how data scientists might use exploratory subgroup analysis to understand a dataset.

\subsection{Formative Design Study Protocol}
\label{app:formative-design}

\noindent \textbf{Introduction.} We’re developing a tool to interactively conduct subgroup or slice analysis. As an expert in \textit{[insert field]} with experience on subgroup analysis, I’d like to learn about how you’ve used these techniques in the past, and if there are ways we can build tools to help you do that more easily.

\noindent \textbf{Part 0: Metadata (5 mins).} 
\begin{itemize}
    \item Could you describe your research or work interests at a very high level?
\end{itemize}

\noindent \textbf{Part 1: Previous Experience with Subgroup Analysis (10–20 mins).}
To ensure we’re on the same page, subgroup analysis seeks to identify data segments (e.g., "female, PhD") where a model’s performance is significantly worse compared to overall performance. Insights from these segments can guide data collection, rules implementation, or model improvement.

\begin{itemize}
    \item How many projects (\textgreater 1 month) have you worked on involving subgroup analysis or similar processes?
    \item Can you describe your most recent project using subgroup analysis?
    \begin{itemize}
        \item What kind of data features did you work with?
        \item Did you need to transform or discretize these features? How did you make those decisions?
        \item What outcomes did you compare within the subgroups?
        \item How did you perform the subgroup analysis? What did you find?
        \item How did the findings help you analyze your data or improve your model?
        \item If any, what were the main obstacles in the project?
    \end{itemize}
    \item Do you typically evaluate on subgroups that you already know about or do you also need to discover new subgroups?
    \begin{itemize}
        \item How do you discover these subgroups? What are the characteristics that would make an interesting subgroup?
    \end{itemize}
    \item Have you used any pre-existing tools, either visual or programmatic, to do subgroup analysis? What tools?
    \begin{itemize}
        \item \textit{(If yes)} Did you find anything particularly helpful, unhelpful, or missing in each tool?
        \item What kinds of visualization did you create or want to create?
    \end{itemize}
    \item Have you ever been surprised by something you found in a subgroup analysis? How did you find it and what was its significance?
    \item What do you do with the results of a subgroup analysis? 
    \begin{itemize}
        \item What is most useful or necessary to communicate when you present these results to others?
    \end{itemize}
\end{itemize}

\noindent \textbf{Part 2: Questions on Divisi Prototypes (25 mins).}
\textit{Interviewer opens a Jupyter Notebook with three prototypes of Divisi loaded and shares their screen.}
Imagine you’re a data scientist working for an airline company, tasked with analyzing passenger satisfaction data to predict satisfaction and identify failure points.

\textbf{Subgroup Results from Divisi Algorithm (No Interface).}
We trained a model to predict the overall satisfaction, and extracted the top 10 results from the subgroup discovery algorithm where the error rate was higher than average.
\begin{itemize}
    \item Interpret the subgroups shown. What types of ratings seem harder for the model to classify?
    \item Is this similar to any analysis you’ve done? If so, how helpful was it? If not, would such an analysis be useful for your data?
    \item How do these subgroups compare to the subgroups you’ve looked at in the past in terms of features and their selection?
    \item If this were part of a tool, what additional features or information would you need to better understand the data?
\end{itemize}

\textbf{Basic Interactive Subgroup Discovery Interface.}
The same 10 subgroups as above are now shown in a subgroup discovery interface with basic re-ranking and interactive search functionalities.
\begin{itemize}
    \item What do you think about this interactive interface?
    \begin{itemize}
        \item Would you find it helpful to re-rank subgroups by different metrics? What metrics might you be interested in?
    \end{itemize}
    \item How understandable are these subgroups? Would you find it helpful to have additional explanations of them, such as using a language model or other summarization technique?
    \item Using this tool as a Jupyter notebook widget, how would you see a tool like this fitting into your workflow?
    \item What additional features or information do you think would be most helpful to better understand the data?
\end{itemize}

\textbf{Visualization: Subgroup Map.}
Finally, we want to show you an additional version of the interface where you are able to interact with a visualization of the dataset in addition to the features available for the last prototype. \textit{Interviewer explains how the Subgroup Map works.}
\begin{itemize}
    \item How would you interpret the overlaps between the subgroups as shown on the plot?
    \item How might this ability to see how subgroups overlap influence your analysis?
    \item What kind of visualization style (for example, dimensionality reduction, Venn diagram) would be more helpful for your analyses? Why?
    \item Do you have other suggestions for improving this interface?
\end{itemize}

\noindent \textbf{Part 3: Final Thoughts (5 mins).}
Let’s reflect on your earlier projects:  
\begin{itemize}
    \item After seeing these tools, how would you envision using subgroup analysis in your workflow?
    \item \textit{(If not interested in subgroup analysis)} Do you see this tool being more helpful for other data science problems?
    \item Which features of the tools did you find the most and least useful?
    \item Are there any analyses you would be interested in performing in the future? What would you need to make those analyses possible?
    \item Is there anything else you’d like to add to what we’ve discussed?
\end{itemize}

\subsection{User Study Protocol}
\label{app:user-study}

\noindent \textbf{Introduction.} Thank you for participating in this study. In this session we’ll have you try out a new tool to discover subgroups in a dataset, and we’re interested in learning how you might use this type of analysis in your work. We appreciate any feedback you can give us!

\noindent \textbf{Part 0: Pre-Study Survey (10 mins).}
\begin{itemize}
    \item Are you familiar with the term subgroup analysis? What does it mean to you? \textit{(If clarification is needed)} Subgroup analysis identifies data segments (e.g., combinations like ``female, PhD degree'') where a model performs poorly.  
    \item Please complete this survey about your prior experience with data science and subgroup analysis. Please share your screen while filling out the survey, and feel free to explain your responses as you go.
    \begin{itemize}
        \item Which of the following best describes your current occupation?
        \item How many years of experience do you have working with data and/or machine learning?
        \item What is your level of comfortability with writing Python code?
        \item How many projects (> 1 month long) have you worked on that have involved subgroup analysis?
        \item Do you typically evaluate on subgroups that you already know about or do you also need to discover new subgroups?
        \item Have you used any pre-existing tools, either visual or programmatic, to do subgroup analysis?
    \end{itemize}
    \item Among your prior projects, which do you think would most benefit from subgroup analysis? Did you use any subgroup analysis for that project?
\end{itemize}

\noindent \textbf{Part 1: First Task (15 mins).}
Imagine you are a data scientist analyzing airline passenger satisfaction data. Your task is to predict customer satisfaction and identify where predictions fail. \textit{[Interviewer walks participant through the dataset using a description shown in the survey form.]}
\begin{itemize}
    \item How would you approach this task using any methods you know?  
\end{itemize}

Please open the tool via the provided link. \textit{[Participant opens Jupyter Notebook interface while sharing screen. Interviewer gives a brief tutorial of how to use the interface, using an annotated screenshot embedded in the notebook.]}

For the first task, we’re just going to be looking at the customers’ overall satisfaction rating. For the next 15 minutes, we’d like you to pretend like you truly are the data scientist at the airline company. It’s your job to find insights in the data about what led to customer dissatisfaction. Please talk aloud as you do your analysis, and explain any insights you come up with. If there’s something you want to do with the tool but aren’t sure how, just ask and we can help.

\textit{Follow-up questions to ask during the study:}
\begin{itemize}
    \item Why did you focus on this particular subgroup?  
    \item What made you decide on the way you have the ranking functions set?  
    \item Would it matter to you if the subgroups you selected represented very similar sets of points? Would you want to check that before you present these subgroups?
    \item Would you want to see how much each feature contributed to the overall rate?  
    \item What would you want to do to make the search results more relevant?
\end{itemize}

\noindent \textbf{Part 2: Second Task (20 mins).}
We can imagine that the data science team at the airline company has now built a classification model that can pretty accurately determine whether a customer is going to be dissatisfied based on their survey ratings. But we’re interested in making this model as accurate as possible, and so the team would like you to look at subgroups where the model error is higher than average. Again, for the next 15 minutes you are truly a data scientist on the team, so please do the task to the best of your ability. \textit{[Use the same set of follow-up questions from Part 1 as needed.]}

\noindent \textbf{Part 3: Final Thoughts (10 mins).}
For our final questions, let’s think back to at the beginning of the session when you mentioned that it could be helpful to analyze subgroups for \textit{[a project that the participant mentioned]}.
\begin{itemize}
    \item Which of your prior projects would most benefit from subgroup analysis?  
    \item Did you do anything similar to the subgroup discovery process we did today in that prior project?
    \item Do you think it would be important to be able to discover new subgroups to analyze the dataset in that project? Or would it be sufficient to use slices that you were already aware of?  
    \item Do you see any other ways that this tool could help you get a different angle on your data?  
    \item Is there anything that the tool would need to do differently to be most helpful to you?
    \item Is there anything else you'd like to add to what we've already discussed?
\end{itemize}

\end{document}